\documentclass[12pt]{elsarticle}  
\usepackage{amssymb,graphicx,amsmath,graphics,xfrac,url,subfigure,booktabs,adjustbox}
\usepackage{caption}

\newcommand\sr{\mathrm{sr}}

\newcommand\sca{\mathrm{S}}
\newcommand\ten{\mathrm{T}}
\newcommand\Pl{\mathrm{Pl}}
\newcommand\ret{\mathrm{ret}}
\newcommand\D{\mathrm{d}}
\newcommand\ele{\mathrm{l}}
\newcommand\ere{\mathrm{r}}
\newcommand\e{\mathrm{e}}

\newcommand\phai{\mathrm{pi}}
\newcommand\ua{\mathrm{ua}}
\newcommand\im{\mathrm{i}}

\newcommand\ex{\mathrm{ex}}
\newcommand\Mpc{\mathrm{Mpc}}

\begin{document}
	
\title{Study of scalar and tensor power spectra in the generalized Starobinsky inflationary model using semiclassical methods}

\author{Clara Rojas}

\address{Yachay Tech University, School of Physical Sciences and Nanotechnology, Hda. San Jos\'e s/n y Proyecto Yachay, 100119, Urcuqu\'i, Ecuador}
\ead{crojas@yachaytech.edu.ec}

\begin{abstract}
In this work we solved the equation of scalar and tensor perturbations for the generalized Starobinsky inflationary model using the improved uniform approximation method and the phase-integral method up to third-order in deviation.  We  compare our results with the  numerical integration.  We have  obtained that  both semiclassical methods reproduce the scalar power spectra $P_{\sca,\ten}$, the scalar spectral index $n_S$,  and the tensor-to-scalar ratio $r$.  Also we present our results in the $(n_S,r)$ plane.

\end{abstract}

\maketitle
	
\section{Introduction}

Inflationary Cosmology arises like a complement of the Big Bang theory. It was proposed in the eighties \cite{guth:1981} to be a solution of the flatness and horizon problem. Inflation also has the property of produces cosmological perturbations. Scalar cosmological perturbations represents the seeds that give origin to the structure formation in our Universe and the anisotropies of the Cosmic Microwave Background Radiation, whereas tensor cosmological perturbations produces primordial gravitational waves \cite{tamayo:2017}. According to  Planck $2018$ results a nonzero tensor amplitude has not been detected, however recent results have imposed an upper limit in the amplitude of tensor modes $r < 0.044$ for the mode $k = 0.05\,  \textnormal{Mpc}^{-1}$  using Planck data in combination with the BICEP/Keck  measurements from 2015  \cite{tristram:2021}, and $r<0.036$ for $k=0.05\, \textnormal{Mpc}^{-1}$ using  BK18 results  \cite{ade:2021}.

In the literature  there are several models of inflation \cite{martin:2014}, and we have to distinguish which of those are supported by observations. The Starobinsky inflationary model \cite{starobinsky:1980} is currently  supported by observations and has been studied in recent works \cite{truman:2021,truman:2020,samart:2019,adam:2019,granada:2019,chowdhury:2019,paliathanasis:2017,diValentino:2017,linde:2014}.

In last years  a generalized version of the Starobinsky inflationary  model also has been caused of a lot of interest \cite{meza:2021,renzi:2020,canko:2020,cheong:2020,fomin:2020,liu:2018,chakravarty:2015,motohashi:2015}, we call it the \textit{generalized Starobinsky inflationary model}. This model depends on a parameter $p$ that is close to the unity, for $p=1$ we recover the Starobinsky inflationary potential. In the literature the parameter $p$ has been constrained to be $1.92 \lessapprox 2p \leq 2$ \cite{motohashi:2015}, and  Renzi \cite{renzi:2020} have found that the parameter $p$ must be in the range $0.962 \leq p \leq 1.016$.  Based in a recent study  \cite{meza:2021} the parameter $p$ is fixed in $p=1.0004$. This parameter $p$ was fixing doing a numerical study of the generalized Starobinsky inflationary model. In order to find the parameter $p$ that fixes with observations, we compute the cosmological parameters sweeping $p$ between $p=0.095$ and $p=1.005$. We  found that the value of $ p = 1.0004$ reproduces the value of $A_\sca$, $n_\sca$ and $r$ is in agreement with the current observational data \cite{meza:2021}.

In this work we studied the perturbations equations into the  generalized Starobinsky inflationary model in three ways: a) doing the numerical integration  mode by mode \cite{meza:2021}, b) using the slow-roll approximation \cite{meza:2021,renzi:2020,motohashi:2015}, and  c) using semiclassical methods:  the  second-order uniform approximation method and the phase integral approximation up to to third-order in deviation.  The slow-roll approximation is the standard technique used in Inflationary Cosmology, where it is considered that the kinetic energy dominates the potential energy. Semiclassical methods have been successfully applied to calculate the scalar and tensor perturbations $u_{k}$ and $v_{k}$, and consequently  the scalar and tensor power spectra $P_{\sca,\ten}$ for several models of inflation \cite{truman:2021,truman:2020,rojas:2012,rojas:2009,rojas:2007c,rojas:2007b,casadio:2006,casadio:2005a, casadio:2005b,casadio:2005c, habib:2005b,martin:2003a,habib:2002}.  Furthermore,  Zhu \textit{et al.} have applied the third-order uniform approximation method to calculate the scalar power spectrum in the $k-$inflation model  \cite{zhu:2014a,zhu:2014b}, this method can improve the accuracy of our results. Once calculate the scalar an tensor power spectra $P_{\sca,\ten}$ is straightforward  calculate the scalar spectral index $n_S$ and the tensor-to-scalar ratio $r$.

The article is structured as follows: In Sec. 2 we present the generalized Starobinsky  inflationary potential. In Sec. 3  we show the basic equations of inflationary cosmology. Sec. 4 is devoted to solve the movement equations of the Universe both into the slow-roll approximation and numerically.  In section 5 we present the equations for scalar and tensor perturbations. Section 6 is devoted to solve the equation of perturbations using numerical calculation,   the second-order slow-roll approximation, and  semiclassical methods. In section 7, we discuss our results.  Finally, in Sec. 8 we present the conclusions of this work.
 
 \section{Generalized Starobinsky inflationary model}
 
The generalized Starobinsky inflationary model comes form of action for $R^{2p}$ inflation in the Einstein frame and is given by \cite{martin:2014,renzi:2020,motohashi:2015}
 
\begin{equation}
\label{gStarobinsky}
V(\phi)= V_0 \,e^{-2 \sqrt{\frac{2}{3}}\phi}\left(e^{\sqrt{\frac{2}{3}}\phi}-1 \right)^{\frac{2p}{2p-1}},
\end{equation}
 with
 
\begin{equation}
V_0=6 \left(\dfrac{2p -1}{4p} \right) M^2 \left(\dfrac{1}{2p}\right)^{\dfrac{1}{2p-1}},
\end{equation}
 where $\phi$ is the scalar field, $p$ is a real number closed to the unity, that means no integer values of $p$.  Based in a previous study we fixed $p=1.0004$ \cite{meza:2021}.  On the other hand $M$ is fixed to normalized the amplitude of the power spectrum to the observable value in $M =1.30 \times 10^{-5}$  \cite{canko:2020}.
 
\begin{figure}[htbp]
 \centering
 \includegraphics[scale=0.35]{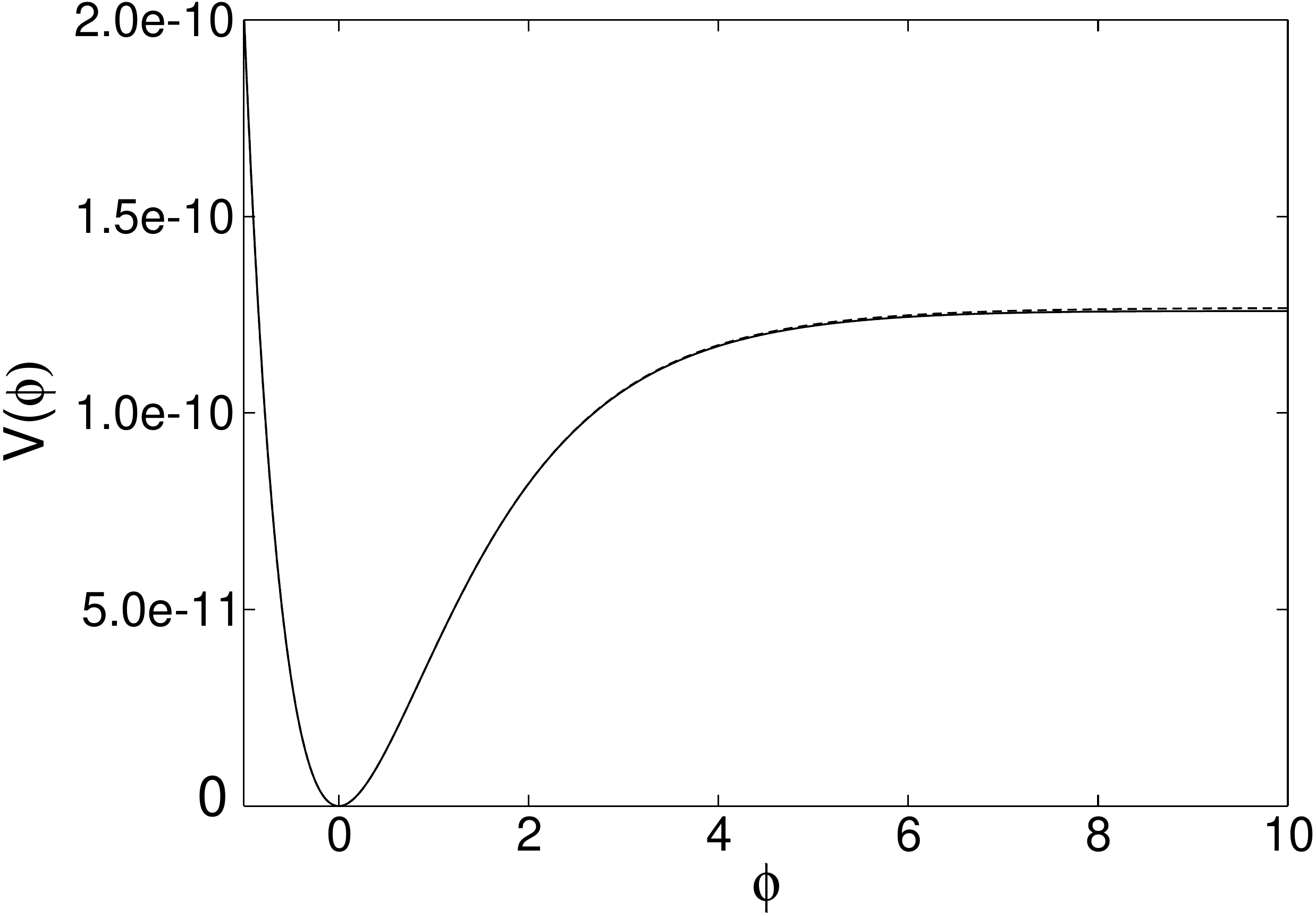}
 \caption{Generalized Starobinsky  inflationary model for $p=1.0004$ (dashed-line), and $p=1$  (solid line), which correspond to the  Starobinsky inflationary model.}
 \label{potential}
 \end{figure}

 At $p=1$, equation \eqref{gStarobinsky} reduces to the original Starobinsky inflationary  model \cite{martin:2014,renzi:2020}. 
 In Fig. \ref{potential} we show the form of the generalized Starobinsky  inflationary model for $p=1.0004$ and $p=1$.
 
 \bigskip

\section{Equations of motion}

The equations of motion of our Universe are given by the Friedmann equation and the fluid equation. Writing the pressure $p$ and the energy density $\rho$ in terms of a scalar field, these equations  are given by \cite{liddle:2000}:

\begin{eqnarray}
\label{Friedmann}
H^2&=&\dfrac{1}{3}\left[V(\phi)+\dfrac{1}{2}\dot{\phi}^2\right],\\
\label{continuity}
\ddot{\phi}&+&3H\dot{\phi}=-V,_\phi,
\end{eqnarray}
where  dots  means derivative respect to the physical time $t$, $V(\phi)$ is the potential of the scalar field, being $V(\phi)$ given by Eq. \eqref{gStarobinsky}, and $V_{,\phi}$ derivatives of the inflationary potential respect to the scalar field $\phi$,
In the generalized Starobinksy inflationary model Eqs. \eqref{Friedmann} and \eqref{continuity} have not exactly solution; they can be solved numerically or using the slow-roll approximation. 

\bigskip
Into the slow-roll approximation \cite{liddle:2000}  we consider that the scalar field  $V(\phi)$ varies very slowly $\dot{\phi}^2 \ll V(\phi)$, then Eqs. \eqref{Friedmann} and \eqref{continuity} reduce to

\begin{eqnarray}
\label{sr2}
H^2 &\simeq& \dfrac{1}{3} V(\phi),\\
\label{sr1}
3 H \dot{\phi} &\simeq& -V_{,\phi}.
\end{eqnarray}

The slow-roll parameters can be expressed in terms of the potential:

\begin{eqnarray}
\epsilon(\phi)&=&\dfrac{1}{2} \left(\dfrac{V'}{V}\right)^2,\\
\eta(\phi)&=& \dfrac{V''}{V}.
\end{eqnarray}

The amount of inflation or the number of e-foldings $N$ is giving by

\begin{equation}
\label{Nex}
N \equiv \dfrac{a(t_{\rm end})}{a(t_{\rm initial})}=\int_{t}^{t_{\rm end}} H \D t.
\end{equation}
 
Into the slow-roll approximation Eq. \eqref{Nex} is giving by

\begin{equation}
\label{N}
N \simeq \int_{\phi}^{\phi_i} \dfrac{V}{V_{,\phi}} \D \phi.
\end{equation}

The amount inflation required to solve the Big Bang problems is about $60 - 70$ e-foldings.

\section{Solutions to the equations of motion}
\subsection{Slow-roll approximation}

In this section we solved the equations of motion into the slow-roll approximation  for the generalized Starobinsky inflationary model. From Eq. \eqref{sr2} we obtain analytically the dependence of the scale factor into the slow-roll approximation $a_\sr$ with the physical time,

\begin{equation}
\dfrac{\D a_\sr}{a_\sr}=\dfrac{1}{\sqrt{3}} \sqrt{V(t)} \, \D t\rightarrow  a_\sr=e^{h(t)},
\end{equation}
where

\vspace{-0.5cm}
\begin{equation}
h(t) = \dfrac{1}{\sqrt{3}} \int_{0}^{t} \sqrt{V(t)} \,\D t.
\end{equation}

From Eq. \eqref{sr1} we obtain numerically the scalar field into the slow-roll approximation $\phi_\sr$,

\begin{equation}
\label{dt}
\D t = -\sqrt{3} \dfrac{\sqrt{V(\phi)}}{V_{,\phi}} \D \phi \rightarrow t =   g(\phi) \rightarrow g(\phi)-t=0,
\end{equation}
where

\vspace{-0.5cm}
\begin{equation}
\label{g}	g(\phi)= -\sqrt{3}\int_{\phi_i}^{\phi}  \dfrac{\sqrt{V(\phi)}}{V_{,\phi}} \D \phi.
\end{equation}

Doing the integration of Eq. \eqref{g} we obtain the function $g(\phi)$ in terms of hypergeometric functions

\vspace{-0.5cm}
\begin{eqnarray}
\nonumber
g(\phi)&=&-\dfrac{3}{M} \dfrac{(-1+2p)^{\sfrac{3}{2}}}{(-2+3p)} 2^{\frac{2-3p}{-1+2p}} p^\frac{1-p}{-1+2p} \\
\nonumber
&\times& \left\{ \left(-1+e^{\sqrt{\frac{2}{3}}\phi}\right)^{\frac{2-3p}{1-2p}}\right.\\
\nonumber
&\times& _{2}F_{1}\left[1,\dfrac{2-3p}{1-2p}, \dfrac{3-5p}{1-2p},\dfrac{(-1+p)}{p} \left(-1+e^{\sqrt{\frac{2}{3}}\phi} \right)\right]\\
\nonumber          
&-& \left(-1+e^{\sqrt{\frac{2}{3}}\phi_i}\right)^{\frac{2-3p}{1-2p}}\\    \nonumber                
&\times& \left. _{2}F_{1}\left[1,\dfrac{2-3p}{1-2p}, \dfrac{3-5p}{1-2p},\dfrac{(-1+p)}{p} \left(-1+e^{\sqrt{\frac{2}{3}}\phi_i} \right)\right] \right\}.\\
\end{eqnarray}

Finally, we solve Eq. \eqref{dt} and obtain $\phi_\sr$ numerically, we call this $
\phi_\sr(p,\phi_i,t)$.


\subsection{Numerical solution}

In this section we solve numerically the complete equations of motion Eqs.  \eqref{Friedmann} and \eqref{continuity}. These equations form a system of coupled differential equations whose solution give us  the behaviour of the scalar field $\phi_\ex(t)$ and the scale factor $a_\ex(t)$, with the physical  time $t$.

Fig. \ref{phisr,asr}(a) shows the evolution of the scalar field $\phi$, we can observed that at $t_{\rm end}=1.02 \times 10^7$ the scalar field starts to oscillate, then inflation ends. Note that  the solution into the slow-roll approximation does not oscillate. 
In Fig. \ref{phisr,asr}(b) we can observed the behaviour of the scale factor $a_\ex(t)$.

\begin{figure}[htp]
\subfigure[]{\label{sr:a}\includegraphics[scale=0.27]{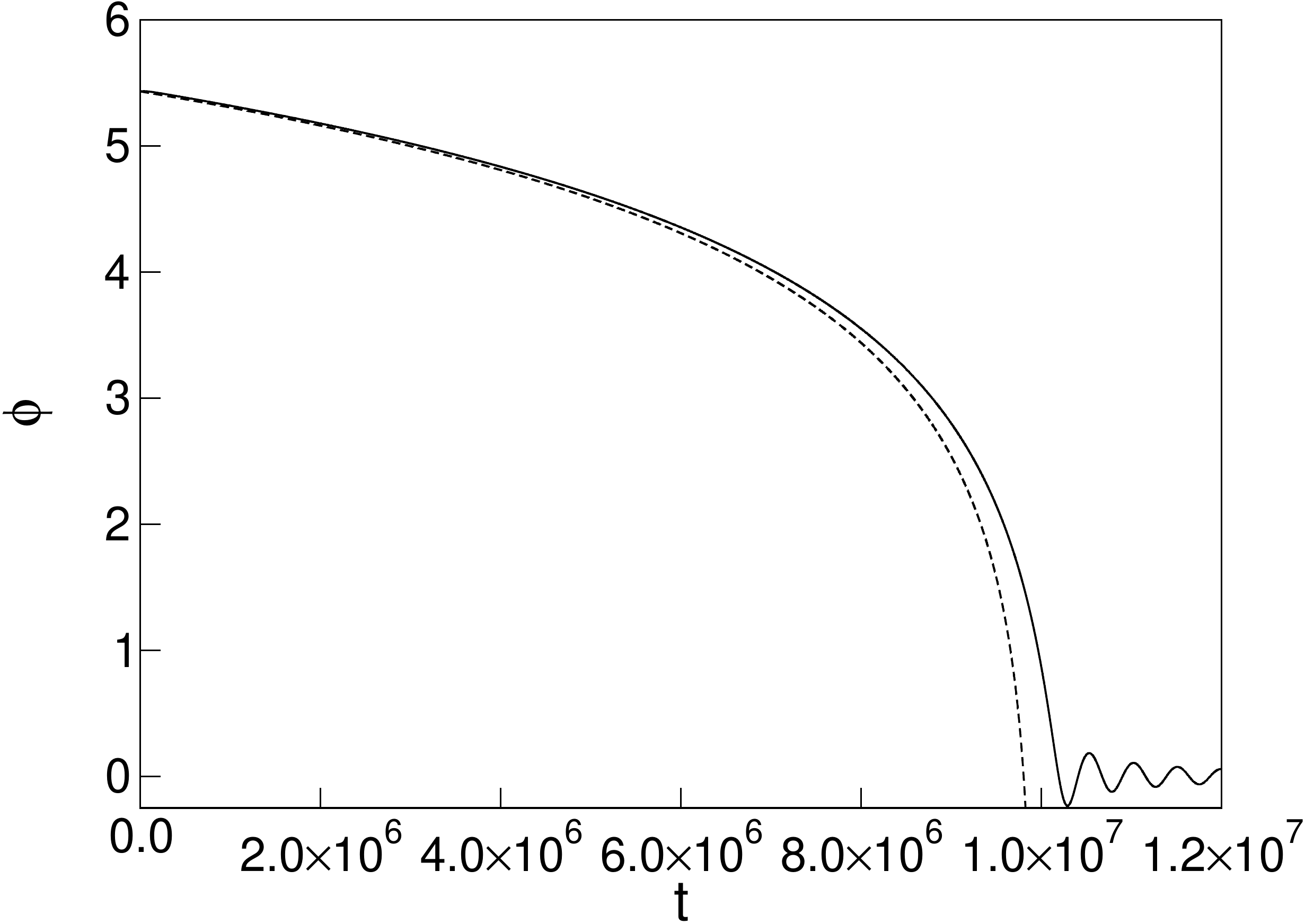}}
\subfigure[]{\label{sr:b}\includegraphics[scale=0.27]{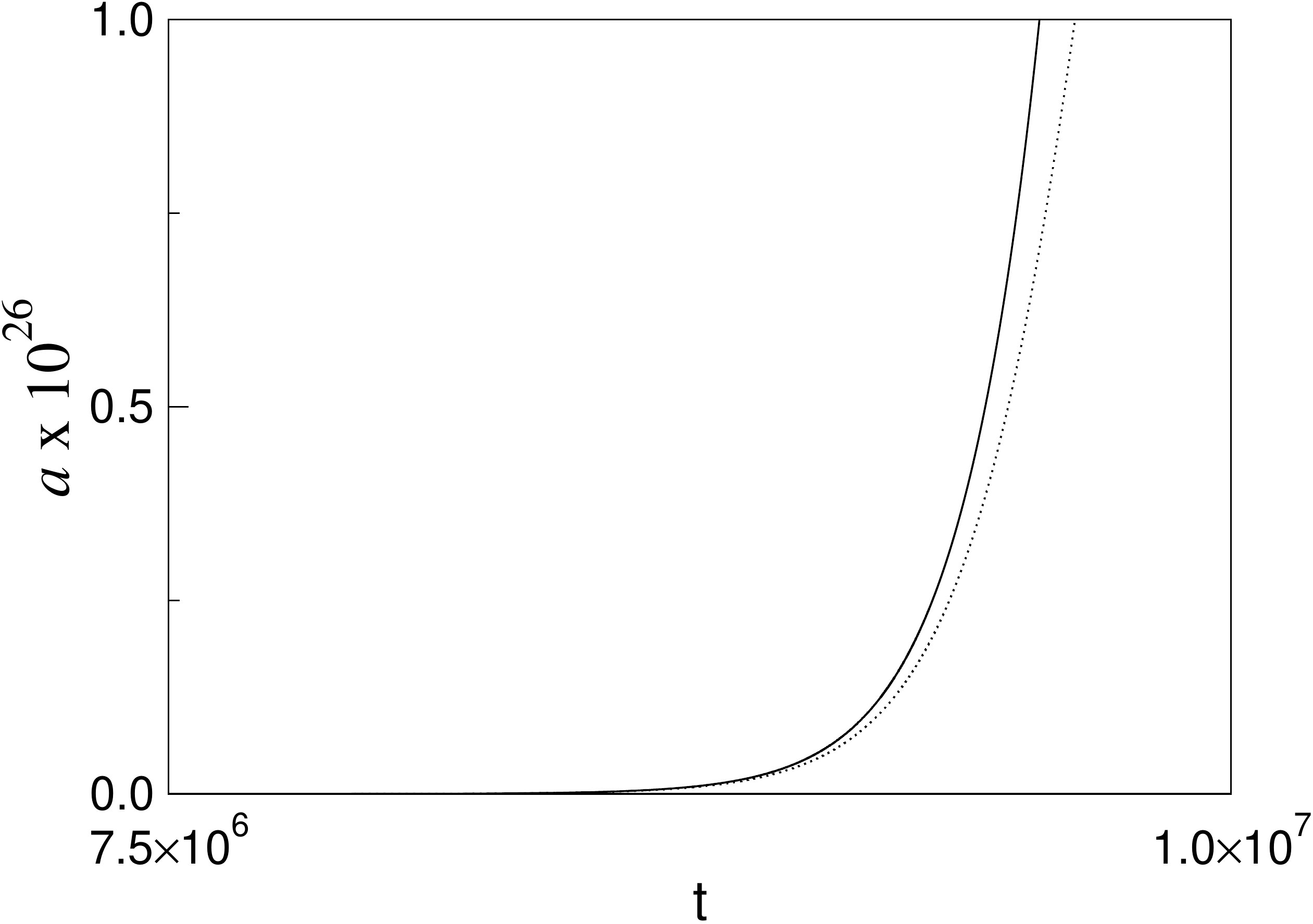}}
\caption{(a) Scalar field $\phi(t)$  and  (b) scale factor $a(t)$  for the generalized Starobinsky inflationary model with $p=1.0004$ where solid line represents the numerical solution, and dashed line the slow-roll approximation.}
\label{phisr,asr}
\end{figure}	

In order to apply semiclassical methods, we need an algebraic expression for $\phi(t)$ and $a(t)$.  We have done a fit from the numerical solution until $t_{\rm end}$ and we found the following dependence respect to the physical time:

\begin{eqnarray}
 \phi_\textnormal{fit}(t)&=& h_0 \log (h_1\, -h_2 M t),\\
 \nonumber
 a_\textnormal{fit}(t) &=&g_0 \,\textnormal{Exp}\left[ -g_1-\dfrac{1}{4} g_2g_3\tanh\left( g_4-\dfrac{g_5 \,t}{g_6}\right)\tanh\left(g_7-\dfrac{g_8\, t}{g_6}\right)\right],\\
 \end{eqnarray}
where  $h_i$'s and $g_i$'s are  well known constants.
Figs.  \ref{phifit,afit}(a) and  \ref{phifit,afit}(b)  show the behaviour of the scalar field $\phi_{\rm fit}$ and the scale factor $a_{\rm fit}$ until $t_{\rm end}$. We can observed that these expressions adjust to the numerical data.

\begin{figure}[htp]
\subfigure[]{\label{fit:a}\includegraphics[scale=0.27]{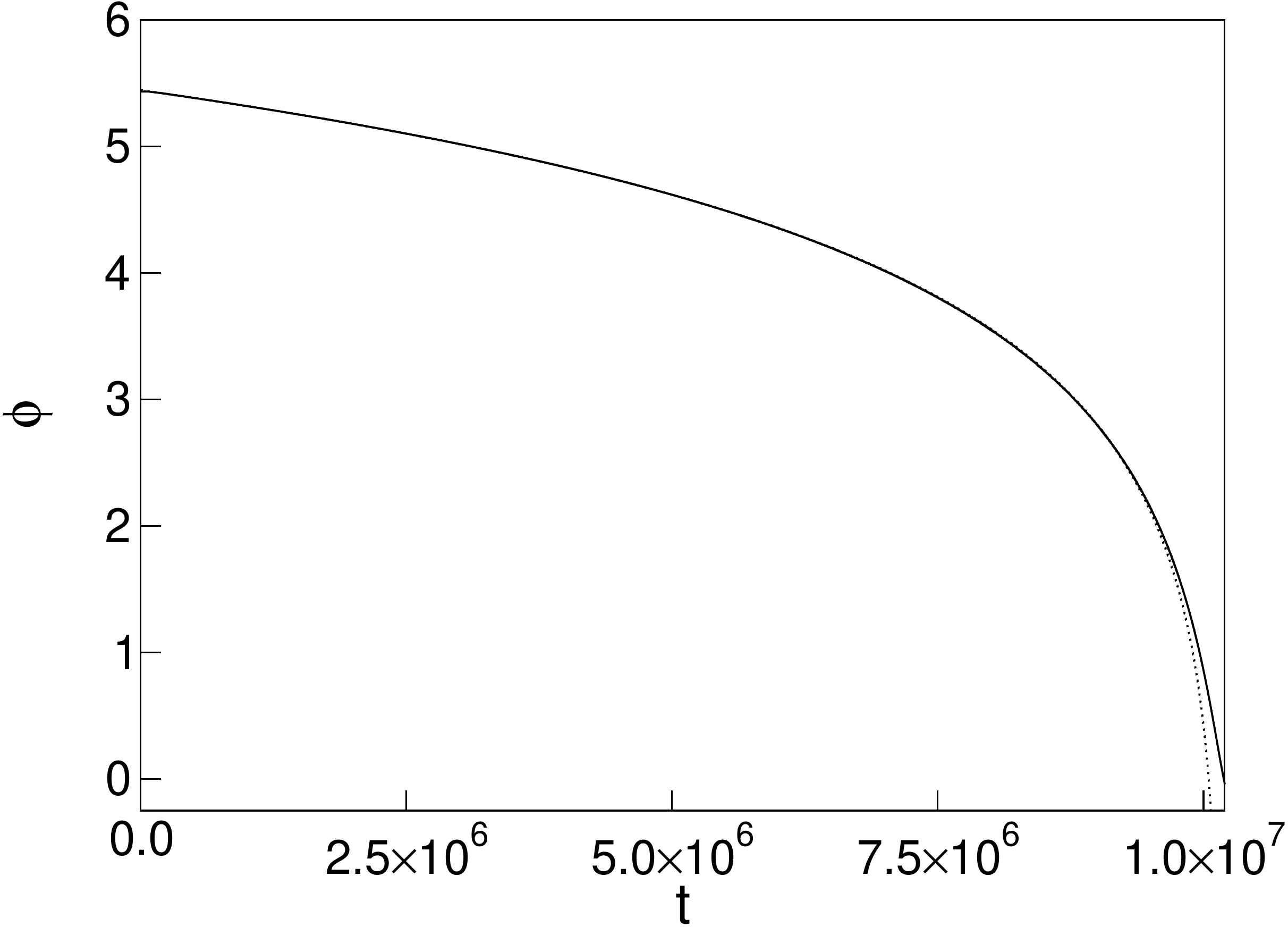}}
\subfigure[]{\label{fit:b}\includegraphics[scale=0.27]{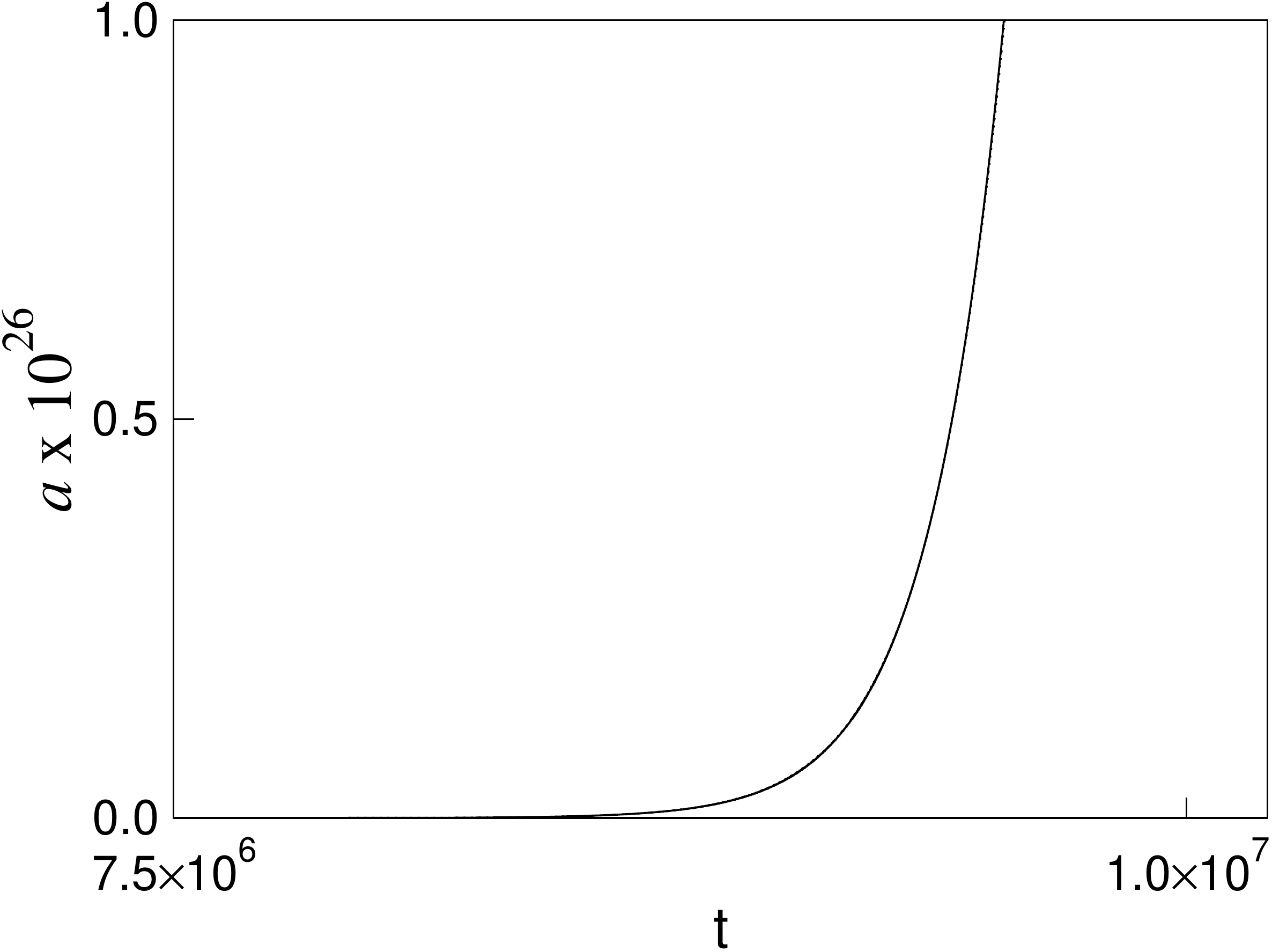}}
\caption{(a) Scalar field $\phi_\textnormal{fit}(t)$  and  (b) scale factor $a_\textnormal{fit}(t)$, respect to the numerical solution,  for the generalized Starobinsky inflationary model with $p=1.0004$ where solid line represents the numerical solution, and dashed line the fitted solution.}
\label{phifit,afit}
\end{figure}
	
\begin{figure}[th!]
\subfigure[]{\label{sr:a}\includegraphics[scale=0.27]{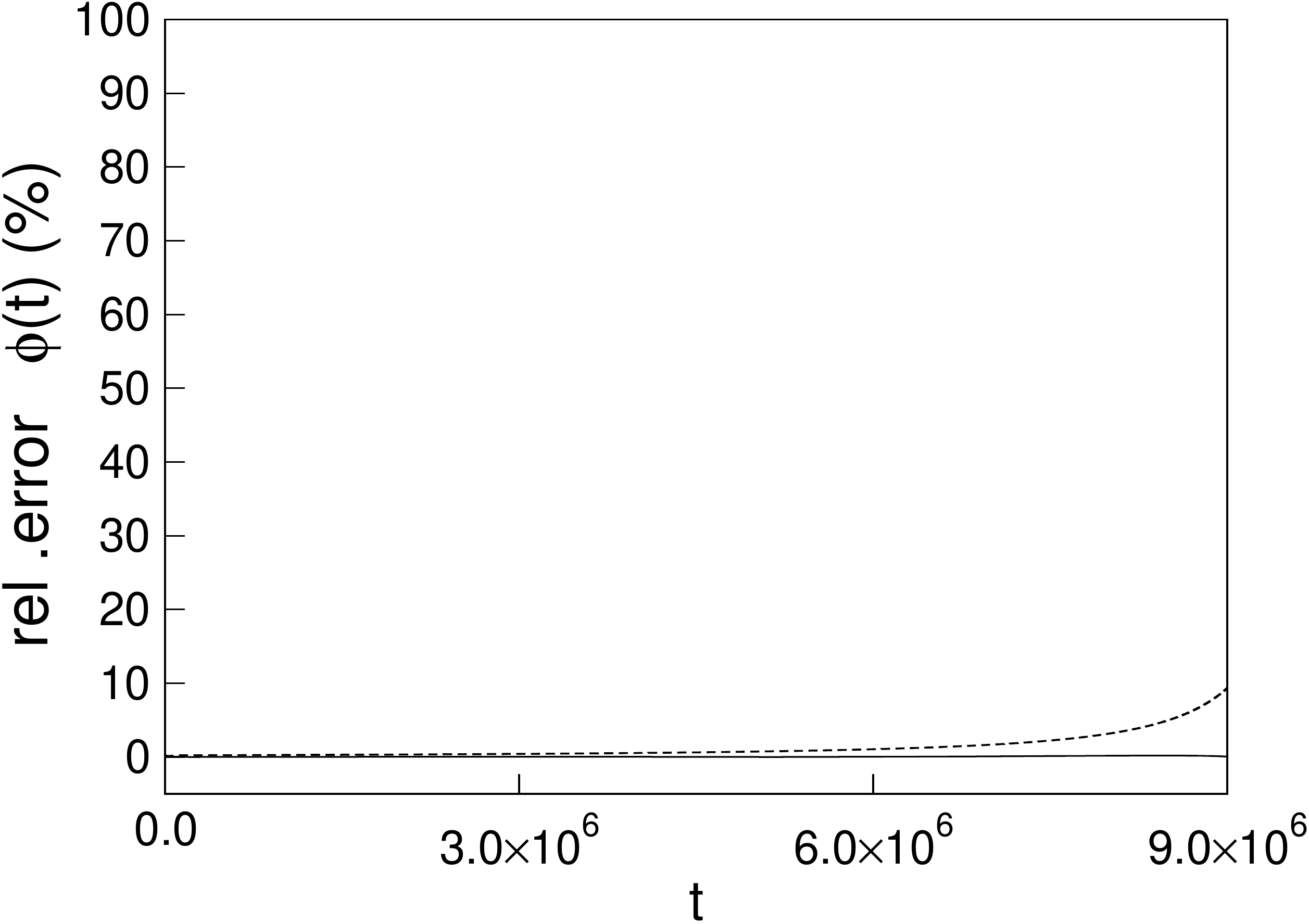}}
\subfigure[]{\label{sr:b}\includegraphics[scale=0.27]{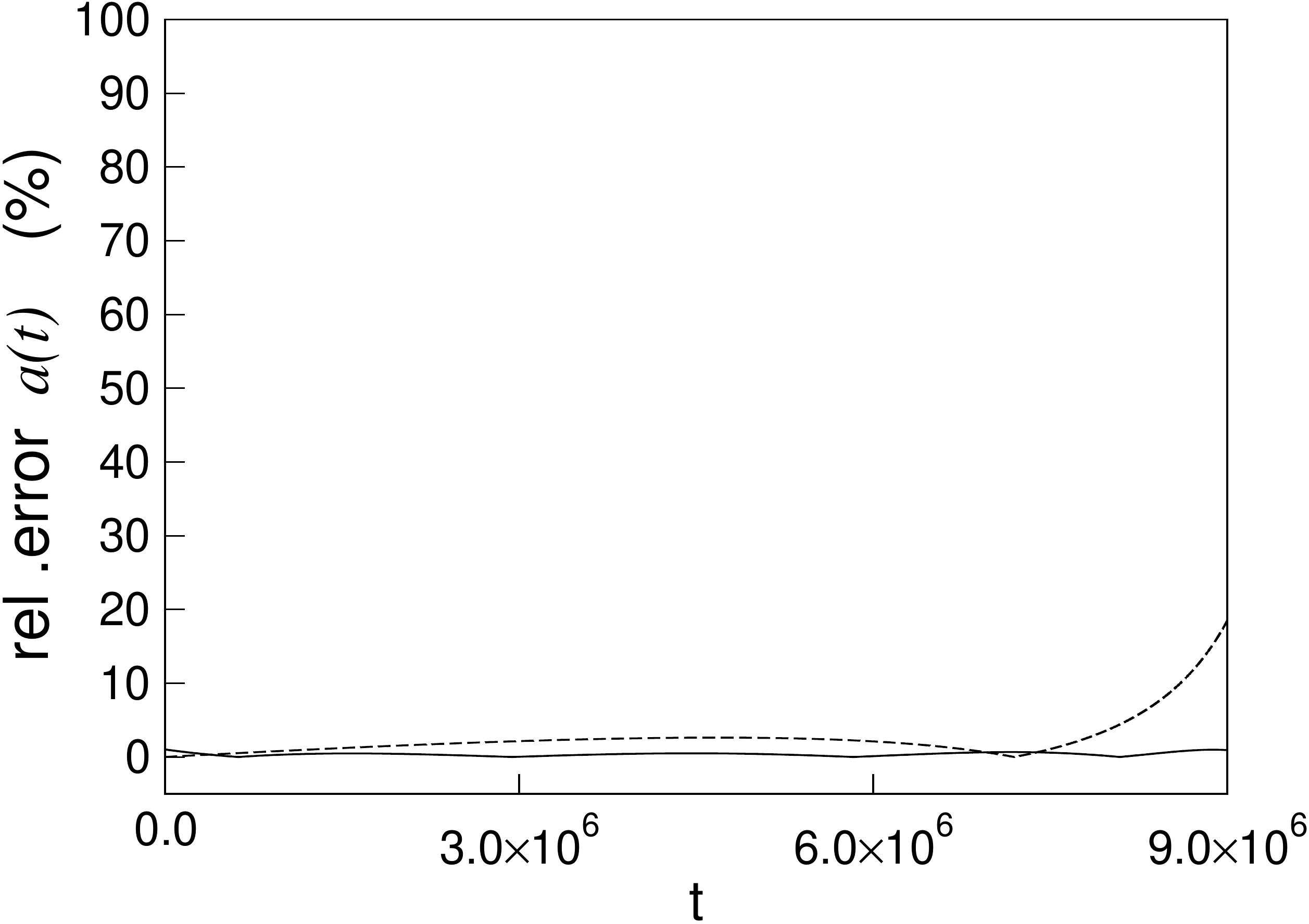}}
\caption{(a) rel. error of the scalar field $\phi(t)$,  and  (b) rel. error of the scale factor $a(t)$  for the generalized Starobinsky inflationary model with $p=1.0004$ where solid line represents the fitted equations, and dashed line the slow-roll approximation.}
\label{error}
\end{figure}

Additionally, we calculate the relative error of the slow-roll approximation and the fitted equations respect to the numerical result.  In Fig, \ref{error}  we can observe that the fitted equations give less relative error than the slow-roll approximation.

\section{Equations for scalar and tensor perturbations}

The scalar perturbations are described by the function $u=a\Phi/\phi'$, where $\Phi$ is a gauge-invariant variable corresponding to the Newtonian potential. The equations of motion of the perturbation $u_k$ in Fourier space are

\begin{equation}
\label{dotdotuk}
u_k''+\left(k^2-\dfrac{z_{S}''}{z_{S}}\right)u_k=0,
\end{equation}
where $z_{S}=a\phi'/\mathcal{H}$, $\mathcal{H}=a'/a$, and the prime indicates derivative with respect to the conformal time $\eta$. The relation between $t$ and $\eta$ is given via the equation $\D t=a\,\D \eta$.

For tensor perturbations one introduces the function $v_k=ah$, where $h$ represents the amplitude of the gravitational
wave. Tensor perturbations obey  a second order differential equation analogous to Eq. (\ref{dotdotuk}):

\begin{equation}
\label{dotdotvk}
v_k''+\left(k^2-\dfrac{a''}{a}\right)v_k=0.
\end{equation}
Considering the limits  $k^2\gg|z_{S}''/z_{S}|$ (short wavelength) and $k^2\ll|z_{S}''/z_{S}|$  (long wavelength), we have that  the  solutions to
Eq. (\ref{dotdotuk}) exhibit the following asymptotic behavior:

\begin{equation}
\label{boundary_0}
u_k\rightarrow \dfrac{e^{-ik\eta}}{\sqrt{2k}}
\quad \left(k^2\gg|z_{S}''/z_{S}|, -k\eta\rightarrow \infty \right),
\end{equation}

\begin{equation}
\label{boundary_i} u_k\rightarrow A_k z  \quad \left(k^2\ll|z_{S}''/z_{S}|,-k\eta\rightarrow 0\right).
\end{equation}

\noindent Equation \eqref{boundary_0} is used as the initial condition for the perturbations. The same asymptotic conditions hold for tensor
perturbations.

The power spectra for scalar and tensor perturbations are given by the expressions

\vspace{-0.5cm}
\begin{eqnarray}
\label{PS}
P_\sca(k)&=& \lim_{kt\rightarrow \infty} \dfrac{k^3}{2 \pi^2}\left|\dfrac{u_k(t)}{z_{S}(t)} \right|^2,\\
\label{PT}
P_\ten(k)&=& \lim_{kt\rightarrow \infty} \dfrac{k^3}{2 \pi^2}\left|\dfrac{v_k(t)}{a(t)} \right|^2,
\end{eqnarray}
and the spectral index for scalar perturbations is defined by:

\begin{equation}
\label{nS}
n_\sca(k)= 1+\dfrac{\D\ln P_\sca(k)}{\D\ln k}.
\end{equation}

\noindent In addition, the tensor-to-scalar ratio $r$ is defined as \cite{habib:2005b}

\begin{equation}
\label{R}
r=8\dfrac{P_\ten(k)}{P_\sca(k)}.
\end{equation}

The scale factor $a$ and the scalar field $\phi$ are obtained  in terms of the physical time $t$ instead the conformal time $\eta$, then we proceed to write the equations for the scalar and tensor perturbations in the variable $t$.  In this case, the equation for the perturbations can be written as

\begin{eqnarray}
\label{dotu}
\ddot{u_k}+\frac{\dot{a}}{a}\dot{u_k}+\frac{1}{a^2}\left[k^2-\frac{\left(\dot{a}\dot{z_\sca}+a\ddot{z_\sca}\right)a}{z_\sca} \right]u_k&=&0,\\
\label{dotv}
\ddot{v_k}+\frac{\dot{a}}{a}\dot{v_k}+\frac{1}{a^2}\left[k^2-\left(\dot{a}^2+a\ddot{a}\right) \right]v_k&=&0.
\end{eqnarray}

\section{Solutions of the perturbation equation}

\subsection{Second-order slow-roll approximation}

The scalar and tensor power spectra in the slow-roll approximation up-to second-order are given by the expressions \cite{stewart:2001}

\begin{eqnarray}
\label{PS_sr}
\nonumber
P_\sca^{\sr}(k)&\simeq&\left[1+(4b-2)\epsilon_1+2b\delta_1+\left(3b^2+2b-22+\frac{29\pi^2}{12}\right)\right.\\
\nonumber
&  +   & \left. 
 \epsilon_1\delta_1+\left(3b^2-4+\frac{5\pi^2}{12}\right)\delta_1^2\left(-b^2+\frac{\pi^2}{12}\right)\delta_2\right]\left(\frac{H}{2\pi}\right)^2\left(\frac{H}{\dot{\phi}}\right)^2\Big.\Big|_{k=aH} ,\\
\\
\nonumber
\label{PT_sr}
P_\ten^{\sr}(k)&\simeq&\Big[1+(2b-2)\epsilon_1+\left(2b^2-2b-3+\dfrac{\pi^2}{2}\right)\epsilon_1^2\\
&+&\left(-b^2+2b-2+\dfrac{\pi^2}{12}\right)\epsilon_2\Big]\left.\left(\dfrac{H}{2\pi}\right)^2\right|_{k=aH},
\end{eqnarray}
where $k=aH$ represents  the horizon crossing,   $b=0.729637$ is the Euler constant, and

\begin{eqnarray}
\epsilon_1&=&-\dfrac{\dot{H}}{H^2},\\
\epsilon_2&=&\dfrac{1}{H}\dfrac{\D\epsilon_1}{\D t},\\
\delta_1&=&\frac{1}{H\dot{\phi}}\frac{\D^2\phi}{\D t^2},\\
\delta_2&=&\frac{1}{H^2\dot{\phi}}\frac{\D^3\phi}{\D t^3}.
\end{eqnarray}

The scalar spectral index and the tensor-to-scalar ratio \cite{casadio:2006} are given by:

\begin{eqnarray}
\label{nS_sr}
 n_\sca^{\sr}(k)&\simeq&1-4\epsilon_1-2\delta_1+(8c-8)\epsilon_1^2+(10c-6)\epsilon_1\delta_1,\\
\label{r}
r & \simeq & 16\, \epsilon_1\left(1+C \epsilon_2\right),
\end{eqnarray}
where $C=-0.7296$.

The expressions  \eqref{PS_sr}, \eqref{PT_sr}, \eqref{nS_sr}, and \eqref{r}  depend explicitly on time. In order to compute these quantities we need to obtain the dependence  on the variable $k$. For a given value of $k$ ($0.0001 \,\Mpc^{-1} \leq k \leq 10 \,\Mpc^{-1}$) we obtain  $t$ from the relation $k=aH$. 

\subsection{Uniform approximation method}
In order to apply semiclassical methods, we eliminate the terms $\dot{u}_k$ and $\dot{v}_k$ in Eq. \eqref{dotu} and Eqs. \eqref{dotv}. We make the change of variables $u_k(t)=\sfrac{U_k(t)}{\sqrt{a}}$ and $v_k(t)=\sfrac{V_k(t)}{\sqrt{a}}$, obtaining that $U_{k}$ and $V_{k}$ satisfy the differential equations:

\begin{eqnarray}
\label{ch2_ddotUk}
\ddot{U}_k+R_\sca(k,t)U_k&=&0,\\
\label{ch2_ddotVk}
\ddot{V}_k+R_\ten(k,t)V_k&=&0,
\end{eqnarray}
with

\begin{eqnarray}
\label{RS}
R_\sca(k,t)&=&\frac{1}{a^2}\left[k^2-\frac{\left(\dot{a}\dot{z_\sca}+a\ddot{z_\sca}\right)a}{z_\sca} \right]+\frac{1}{4a^2}\left(a^2-2a\ddot{a}\right),\\
\label{RT}	R_\ten(k,t)&=&\frac{1}{a^2}\left[k^2-\left(\dot{a}^2+a\ddot{a}\right) \right]+\frac{1}{4a^2}\left(a^2-2a\ddot{a}\right),
\end{eqnarray}
where $U(k)$ satisfies the asymptotic conditions

\begin{eqnarray}
\label{ch2_cero_Uk}
U_k&\rightarrow&A_k \sqrt{a(t)} z_\sca(t),\quad  k\,t\rightarrow \infty,\\
\label{ch2_borde_Uk}
U_k&\rightarrow&\sqrt{\frac{a(t)}{2k}}\exp{\left[-ik\eta(t)\right]}, \quad k\,t\rightarrow 0,
\end{eqnarray}
the asymptotic conditions (\ref{ch2_cero_Uk}) and (\ref{ch2_borde_Uk}) also hold for $V_k$.

We want to obtain an approximate solution of the differential equation \eqref{ch2_ddotUk} and  \eqref{ch2_ddotVk}  in terms of the known solutions $w_\sca(\rho_\sca)$ and $w_\ten(\rho_\ten)$ of the comparison equation \cite{berry:1972,habib:2002,rojas:2007b,rojas:2007c}: 

\begin{eqnarray}
\label{comparison_Sca}
\frac{\D ^2 w_\sca(\rho_\sca)}{\D \rho_\sca^2}+r_\sca(\rho_\sca) w_\sca(\rho_\sca)=0,\\
\label{comparison_Ten}
\frac{\D ^2 w_\ten(\rho_\ten)}{\D \rho_\ten^2}+r_\ten(\rho_\ten) w_\ten(\rho)=0,
\end{eqnarray}
where $r_\sca(\rho_\sca)$ is chosen similar to $R_\sca(k, t)$ and $r_\ten(\rho_\ten)$ is chosen similar to $R_\ten(k, t)$,  with the same number of zeros, so that the solutions of equations \eqref{comparison_Sca} and \eqref{comparison_Ten} are known. 

The functions $U(k,t)$ and $V(k,t)$ must also be similar to $w_\sca(\rho_\sca)$ and $w_\ten(\rho_\ten)$, they can be related via \cite{berry:1972} 

\begin{eqnarray}
\label{relationS}
U_k(k,t) \approx \left\{\dfrac{r_\sca\left[\rho_\sca(k,t)\right]}{R_\sca(k,t)} \right\}^{\sfrac{1}{4}} w_\sca\left[\rho_\sca(k,t)\right],\\
\label{relationT}
V_k(k,t) \approx \left\{\dfrac{r_\ten\left[\rho_\ten(k,t)\right]}{R_\ten(k,t)} \right\}^{\sfrac{1}{4}} w_\ten\left[\rho_\ten(k,t)\right].
\end{eqnarray}

The validity condition to Eqs. \eqref{relationS} and \eqref{relationT} be a good solution is given by

\begin{equation}
\left|\dfrac{1}{R_{\sca,\ten}(t)} \left(\dfrac{\D \rho_{\sca,\ten}}{\D t} \right)\dfrac{\D^2}{\D t^2}\left(\dfrac{\D \rho_{\sca,\ten}}{\D t} \right)^{\sfrac{1}{2}}  \right|\ll 1.
\end{equation}

Eqs. \eqref{relationS} and \eqref{relationT} give an uniform approximation for $U_k(k,t)$ and $V_k(k,t)$ for the complete range of $t$, including the turning points.

To find an approximate solution to the differential equations \eqref{ch2_ddotUk} and \eqref{ch2_ddotVk} in a region where  $Q_\sca^2(k,t)$ and   $Q_\ten^2(k,t)$ have a simple root  at $t_\ret=\tau_\sca$, and $t_\ret=\tau_\ten$, respectively, so that $Q_{\sca,\ten}^2(k,t)>0$ for  $0<t<t_\ret$ and $Q_{\sca,\ten}^2(k,t)<0$ for  $t>t_\ret$ as depicted in Fig. \ref{ch2:QSa} and Fig. \ref{ch2:QTa}.  A suitable comparison function is $ r_\sca(\rho)=\pm\rho_\sca  $ and $r_\ten(\rho)=\pm\rho_\ten$, therefore there are two cases: 

\begin{description}
\item{a)} In the classically allowed region, $Q_{\sca,\ten}^2(k, t)> 0$,  we choose $r_{\sca,\ten}(\rho_{\sca,\ten}) = \rho_{\sca,\ten}$ and the comparison equations to solved are 
	
\begin{eqnarray}
\label{eqr,rk=rho_Sca}
\frac{\D ^2w_\sca}{\D \rho_\sca^2}+\rho_\sca\, w_\sca=0,\\
\label{eqr,rk=rho_Ten}
\frac{\D ^2w_\ten}{\D \rho_\ten^2}+\rho_\ten w_\ten=0.
\end{eqnarray}
	
Eq. \eqref{eqr,rk=rho_Sca}  and \eqref{eqr,rk=rho_Ten} are the Airy equation that has two independent solutions $A_i(-\rho_{\sca,\ten}) $ and $B_i(-\rho_{\sca,\ten})$ \cite{abramowitz:1965}. The mapping relation is given by \cite{berry:1972} 
	
\begin{equation}
\label{mapeo,rk=rho}
\frac{\D \rho_{\sca,\ten}}{\D t}=\left[\frac{Q_{\sca,\ten}^2(k,t)}{\rho_{\sca,\ten}}\right]^{1/2}.
\end{equation}
	
Finally, the approximate solutions to the differential equations \eqref{ch2_ddotUk} and \eqref{ch2_ddotVk}  are

\begin{eqnarray}
\label{Uk_zero}
\nonumber
U_k(k,t)&=&\left[\frac{\rho^\ele_{\sca}(k,t)}{Q_\sca^2(k,t)} \right]^{1/4} \left\{C_1
A_i[-\rho^\ele_{\sca}(k,t)]+C_2 B_i[-\rho^\ele_{\sca}(k,t)] \right\},\\
\\
\label{Vk_zero}
\nonumber
V_k(k,t)&=&\left[\frac{\rho^\ele_{\ten}(k,t)}{Q_\ten^2(k,t)} \right]^{1/4} \left\{C_1
A_i[-\rho^\ele_{\ten}(k,t)]+C_2 B_i[-\rho^\ele_{\ten}(k,t)] \right\},\\
\\
\frac{2}{3}\left[\rho^\ele_{\sca,\ten}(k,t)\right]^{3/2}&=&\int_{t}^{t_\ret} \left[Q_{\sca,\ten}^2(k,t)\right]^{1/2}\D t,
\end{eqnarray}
\medskip
where  $C_1$ and  $C_2$ are two constants to be determined with the help of the boundary conditions \eqref{ch2_borde_Uk}.  
In the limit $kt \rightarrow \infty$, the asymptotic formulas are used  \cite{abramowitz:1965}

\begin{eqnarray}
\label{Ai_1}
A_i(-\rho)&\sim&\pi^{-1/2}\rho^{-1/4}\sin\left(\frac{2}{3}\rho^{3/2}+\frac{\pi}{4}\right),\\
B_i(-\rho)&\sim&\pi^{-1/2}\rho^{-1/4}\cos\left(\frac{2}{3}\rho^{3/2}+\frac{\pi}{4}\right).
\label{Bi_1}
\end{eqnarray}

It is found that $C_1=\sqrt{\frac{\pi}{2}}\e^{-i\pi/4}$ and $C_2=\sqrt{\frac{\pi}{2}}\e^{i\pi/4}$.

\item{b)} In the classically forbiddend region, $Q_{\sca,\ten}^2(k, t)< 0$, we choose $r_{\sca,\ten}(\rho_{\sca,\ten}) = -\rho_{\sca,\ten}$,    and solve the comparison equations

\begin{eqnarray}
\label{eqr,rk=rho_Sca_right}
\frac{\D ^2w_\sca}{\D \rho_\sca^2}-\rho_\sca\, w_\sca=0,\\
\label{eqr,rk=rho_Ten_right}
\frac{\D ^2w_\ten}{\D \rho_\ten^2}-\rho_\ten\, w_\ten=0.
\end{eqnarray}

Eq. \eqref{eqr,rk=rho_Sca_right}  and \eqref{eqr,rk=rho_Ten_right} has the form of the Airy differential equation, which has two independent solutions $A_i(-\rho_{\sca,\ten}) $ and $B_i(-\rho_{\sca,\ten}) $ \cite{abramowitz:1965}. The mapping relation is given by \cite{berry:1972} 

\begin{equation}
\label{mapeo,rk=rho_right}
\frac{\D \rho_{\sca,\ten}}{\D t}=\left[\frac{-Q_{\sca,\ten}^2(k,t)}{\rho_{\sca,\ten}}\right]^{1/2}.
\end{equation}

The approximate solutions to the differential equations \eqref{ch2_ddotUk} and \eqref{ch2_ddotVk}  are

\begin{eqnarray}
\label{Uk_infinity}	
\nonumber
U_k(k,t)&=&\left[\frac{-\rho^\ere_\sca(k,t)}{Q_\sca^2(k,t)} \right]^{1/4} \left\{C_1
A_i[\rho^\ere_\sca(k,t)]+C_2 B_i[\rho^\ere_\sca(k,t)] \right\},\\
\\
\label{Vk_infinity}
\nonumber
V_k(k,t)&=&\left[\frac{-\rho^\ere_\ten(k,t)}{Q_\ten^2(k,t)} \right]^{1/4} \left\{C_1
A_i[\rho^\ere_\ten(k,t)]+C_2 B_i[\rho^\ere_\ten(k,t)] \right\},\\
\\
\frac{2}{3}\left[\rho^\ere_{\sca,\ten}(k,t)\right]^{3/2}&=&\int_{t_\ret}^{t} \left[-Q_{\sca,\ten}^2(k,t)\right]^{1/2}\D t,
\end{eqnarray}

For the computation of the power spectrum we need to take the limit $k\,t\rightarrow \infty$ of the solutions  \eqref{Uk_infinity} and \eqref{Vk_infinity}. In this limit we have

\begin{eqnarray}
\label{Ai_right}
A_i(\rho)&\sim&\dfrac{\pi^{-1/2}}{2}\rho^{-1/4}\exp\left(-\frac{2}{3}\rho^{3/2}\right),\\
B_i(\rho)&\sim&\pi^{-1/2}\rho^{-1/4}\exp\left(\frac{2}{3}\rho^{3/2}\right).
\label{Bi_right}
\end{eqnarray}

Finally,

\begin{eqnarray}
\label{limit_uk}
\nonumber
u_k^\ua(t)&\rightarrow&  \frac{C}{\sqrt{2\,a(t)}}\left[-Q_\sca^2(k,t)\right]^{-1/2}\left\{ \frac{1}{2}\exp\left(-\int_{\tau_\sca}^{t}\left[-Q_\sca^2(k,t)\right]^{1/2} \D t\right)\right.\\
& +&\left.\im\,\exp\left(\int_{\tau_\sca}^{t}\left[-Q_\sca^2(k,t)\right]^{1/2} \D t\right)\right\}\\
\nonumber
\label{limit_vk}
v_k^\ua(t)&\rightarrow&  \frac{C}{\sqrt{2\,a(t)}}\left[-Q_\ten^2(k,t)\right]^{-1/2}\left\{ \frac{1}{2}\exp\left(-\int_{\tau_\ten}^{t}\left[-Q_\ten^2(k,t)\right]^{1/2} \D t\right)\right.\\
& +&\left.\im\,\exp\left(\int_{\tau_\ten}^{t}\left[-Q_\ten^2(k,t)\right]^{1/2} \D t\right)\right\},
\end{eqnarray}
where $C$ is a phase factor.  Using the growing part  of the solutions   \eqref{limit_uk} and  \eqref{limit_vk}   one can compute the scalar and tensor power spectra using the uniform approximation method,

\end{description}
\begin{eqnarray}	
\label{PS_ua}
P_\sca(k)&=&\lim_{-k t\rightarrow \infty} \frac{k^3}{2\pi^2} \left|\frac{u_k^\ua(t)}{z_\sca(t)}\right|^2,\\
\label{PT_ua}
P_\ten(k)&=&\lim_{-k t\rightarrow \infty} \frac{k^3}{2\pi^2} \left|\frac{v_k^\ua(t)}{a(t)}\right|^2.
\end{eqnarray} 
We  use the improved uniform approximation for the calculation of the power spectra \cite{habib:2005b},

\begin{equation}
\label{PS,PT}
\tilde{P}_{\sca,\ten}(k)=P_{\sca,\ten}(k)\left[\Gamma^*(\bar{\tau}_{\sca,\ten})\right]²,
\end{equation}
where $\bar{\tau}_{\sca,\ten}$ is the turning point for the scalar or tensor power spectra and 

\begin{equation}
\Gamma^*(\bar{\tau}_{\sca,\ten})\equiv 1+\frac{1}{12\,\bar{\tau}_{\sca,\ten}}+\frac{1}{288\,\bar{\tau}_{\sca,\ten}^2}-\frac{139}{51840\,\bar{\tau}_{\sca,\ten}^3}+\cdots.
\end{equation}


\subsubsection{Phase-integral method} 
Let us consider the differential equation

\begin{equation}\label{eq_ori}
\frac{d^2u_k}{dz^2}+R(z) u_k=0.
\end{equation}
where  $R(z)$ is an analytic function of  $z$. In order to obtain an approximate solution to Eq. (\ref{eq_ori}), we are going to use the phase-integral   method  developed by Fr\"oman \cite{froman:1965,froman:1966B}. The phase integral approximation, generated using a non specified base solution  $Q(z)$,  is a linear
combination of the phase integral functions \cite{froman:1974A,froman:1996}, which exhibit the following  form

\begin{equation}\label{pi_uk}
u_k=q^{-1/2}(z) \exp\left[\pm i\, \omega(z) \right],
\end{equation}
where

\begin{equation}\label{omega}
\omega(z)=\int^z q(z) dz.
\end{equation}
Substituting (\ref{pi_uk}) into (\ref{eq_ori}) we obtain that the exact phase integrand $q(z)$ must be a solution of the differential equation

\begin{equation}\label{q_ori}	q^{-3/2}(z)\frac{d^2}{dz^2}q^{-1/2}(z)+\frac{R(z)}{q^2(z)}-1=0.
\end{equation}
For any solution of Eq. (\ref{q_ori}) the functions (\ref{pi_uk}), are linearly independent, the linear combination of the functions $u_k$ represents a local solution. In order to solve the global problem we choose a linear combination of phase integral solutions representing the same solution in different regions of the complex plane. This is known as the Stokes phenomenon \cite{froman:1965}.

If we have a function $Q(z)$ which is an approximate solution of Eq. (\ref{q_ori}), the quantity $\epsilon_0$, obtained after substituting $Q(z)$ into Eq. (\ref{q_ori})

\begin{equation}\label{epsilon_0}
\epsilon_0=Q^{-3/2}(z)\frac{d^2}{dz^2}Q^{-1/2}(z)+\frac{R(z)-Q^2(z)}{Q^2(z)},
\end{equation}
is small compared to unity. We take into account the relative small size of $\epsilon_0$ by considering it proportional to $\lambda^2$, where $\lambda$ is a small parameter. The parameter $\epsilon_0$ is small when $Q(z)$ is proportional to $1/\lambda$ and $R(z)-Q^2(z)$ is independent of $\lambda$, i.e. if  $R(z)$ is replaced by $Q^2(z)/\lambda^2+\left[R(z)-Q^2(z)\right]$ in Eq. (\ref{eq_ori}). Therefore, instead of considering Eq. (\ref{eq_ori}), we deal with the auxiliary differential equation

\begin{equation}\label{eq_aux}
\frac{d^2u_k}{dz^2}+\left\{\frac{Q^2(z)}{\lambda^2}+\left[R(z)-Q^2(z)\right]\right\}
u_k=0,
\end{equation}
which reduces to Eq. (\ref{eq_ori}) when  $\lambda=1$.
Inserting the solutions (\ref{pi_uk}) into the auxiliary
differential equation (\ref{eq_aux}), we obtain the following
equation for $q(z)$

\begin{equation}
q^{1/2}\frac{d^2}{dz^2}q^{-1/2}-q^2+\frac{Q^2(z)}{\lambda^2}+R(z)-Q^2(z)=0,
\end{equation}
which is called the auxiliary $q$ equation.  After introducing the
new variable $\xi$,
\begin{equation}
\xi=\int^zQ(z)dz,
\end{equation}
we obtain
\begin{equation}\label{qQdif}
1-\left[\frac{q\lambda}{Q(z)}\right]^2+\epsilon_0\lambda^2+\left[\frac{q\lambda}{Q(z)}\right]^{1/2}\frac{d^2}{d\xi^2}\left[\frac{q\lambda}{Q(z)}\right]^{-1/2}\lambda^2=0,
\end{equation}
where $\epsilon_0$ is defined by Eq. (\ref{epsilon_0}). A formal
solution of Eq. (\ref{qQdif}) is obtained after the identification

\begin{equation}\label{qlambdaQ}
\frac{q\lambda}{Q}=\sum^\infty_{n=0} Y_{2n}\lambda^{2n}.
\end{equation}
Substituting Eq. (\ref{qlambdaQ}) into Eq. (\ref{qQdif}), we obtain

\begin{equation}\label{recurrence}
1-\left(\sum_n
Y_{2n}\lambda^{2n}\right)^2+\epsilon_0\lambda^2+\left(\sum_n
Y_{2n}\lambda^{2n}\right)^{1/2}\frac{d^2}{d\xi^2}\left(\sum_n
Y_{2n}\lambda^{2n}\right)^{-1/2}=0.
\end{equation}

\vspace{0.25cm}
Using computer manipulation algebra it is straightforward to obtain
the coefficients $Y_{2n}$. The first values are
\cite{froman:1966B,campbell:1972}

\vspace{-0.25cm}
\begin{eqnarray}
\label{Y0}
Y_0&=&1,\\
\label{Y2}
Y_2&=&\frac 1 2 \epsilon_0,\\
Y_4&=&-\frac 1 8 \left(\epsilon_0^2+\epsilon_2 \right),\\
Y_6&=&\frac 1 {32} \left(2 \epsilon_0^2 + 6 \epsilon_0\epsilon_2+5\epsilon_1^2+\epsilon_4\right),\\
\end{eqnarray}
where  $\epsilon_\nu$ is defined as

\begin{equation}\label{epsilon_nu}
\epsilon_\nu=\frac{1}{Q(z)}\frac{d\epsilon_{\nu-1}}{dz}, \quad \nu
\ge 1.
\end{equation}
Truncating the series (\ref{qlambdaQ}) at $n=N$ with $\lambda=1$ we
obtain

\begin{equation}\label{qQ}
q(z)=Q(z)\sum^N_{n=0} Y_{2n},
\end{equation}
Substituting (\ref{qQ}) in (\ref{omega}) we have that

\begin{equation}\label{omegadef}
\omega(z)=\sum_{n=0}^N \omega_{2n}(z),
\end{equation}
where

\begin{equation}\label{omega_sum}
\omega_{2n}(z)=\int^z Y_{2n}Q(z) dz.
\end{equation}

From   (\ref{pi_uk}), (\ref{qQ}), and (\ref{omegadef}) we obtain a phase integral approximation of order $2N+1$ generated with the help of the base function $Q(z)$.

The base function $Q(z)$ is not specified and its selection depends on the problem  in question. In many cases, it is enough to choose $Q^2(z)=R(z)$, and the first-order phase integral approximation reduces to the WKB approximation. In the first-order approximation it is convenient to choose a root of $Q^2(z)$ as the lower integration limit in expression (\ref{omega_sum}). However,  for higher orders, i.e. for $2N+1>1$, this is not possible because the function $q(z)$ is singular at the zeros of $Q^2(z)$. In this case, it is convenient to express $\omega_{2n}(z)$ as a contour integral over a two-sheet Riemann surface where $q(z)$ is single valued \cite{froman:1966B}. We define

\begin{equation}
\omega_{2n}(z)=\frac{1}{2} \int_{\Gamma_{t}}Y_{2n}(z)Q(z)dz,
\end{equation}
where  $t$ is a zero of $Q^2(z)$ and  $\Gamma_{t}$ is an integration contour  starting  at the point corresponding to $z$ over a Riemann sheet adjacent to the complex plane, and that encloses the point $t$, in the positive or negative sense and ends at the point $z$.

If the function $Q(z)$ is chosen conveniently, the quantity  $\mu$ defined by

\begin{equation}
\label{mu}\mu=\mu(z,z_0)=\left|\int_{z_0}^z\left|\epsilon(z)q(z)
dz\right|\right|,
\end{equation}
is much smaller than 1.  The function  $\epsilon(z)$ is given by the left side of Eq. (\ref{q_ori})

\begin{equation}
\epsilon(z)=q^{-3/2}(z)\frac{d^2}{dz^2}q^{-1/2}(z)+\frac{R(z)}{q^2(z)}-1,
\end{equation}
where the integral $\mu$ measures  the accuracy of the phase-integral approximation \cite{froman:2002}.

We assume that the function $Q^2(z)$ is real over the real axis. Taking into account this restriction, we shall call turning point, the zero of $Q^2(z)$. We want to know the connection formulas at both sides of an isolated turning point $z_{ret}$, i.e., a turning point which is located far from other turning points. We will adopt the terms  ``classically permitted region'' and  ``classically forbidden region'' in order to denote those ranges over the real axis where $Q^2(z)>0$ and $Q^2(z)<0$, respectively.

The connection formula for an approximate solution that crosses the turning point $z_{ret}$ from a classically permitted region to a classically forbidden region is \cite{froman:1970A}

\begin{equation}\label{allowed-forbbiden}
\left|
q^{-1/2}(z)\right|\cos\left(\left|\omega(z)\right|+\frac{\pi}{4}\right)\rightarrow
\left| q^{-1/2}(z)\right|\exp\left[\left|\omega(z)\right|\right].
\end{equation}

The connection formula for an approximate solution that crosses the turning point $z_{ret}$ from a classically forbidden region to a classically permitted region is \cite{froman:1970A}

\begin{equation}\label{forbbiden-allowed}
\left| q^{-1/2}(z)\right|\exp\left[-\left|\omega(z)\right|\right]
\rightarrow 2 \left|
q^{-1/2}(z)\right|\cos\left(\left|\omega(z)\right|-\frac{\pi}{4}\right).
\end{equation}

It is important to emphasize the one-directional character of the connection formulas (\ref{allowed-forbbiden}) and (\ref{forbbiden-allowed}), this means that  the trace of the solution should be done in the direction indicated by the arrows in Eq. (\ref{allowed-forbbiden}) and Eq. (\ref{forbbiden-allowed}).

\begin{figure}[htbp]
\begin{center}
\subfigure[]{
\label{ch2:QSa}
\includegraphics[scale=0.4]{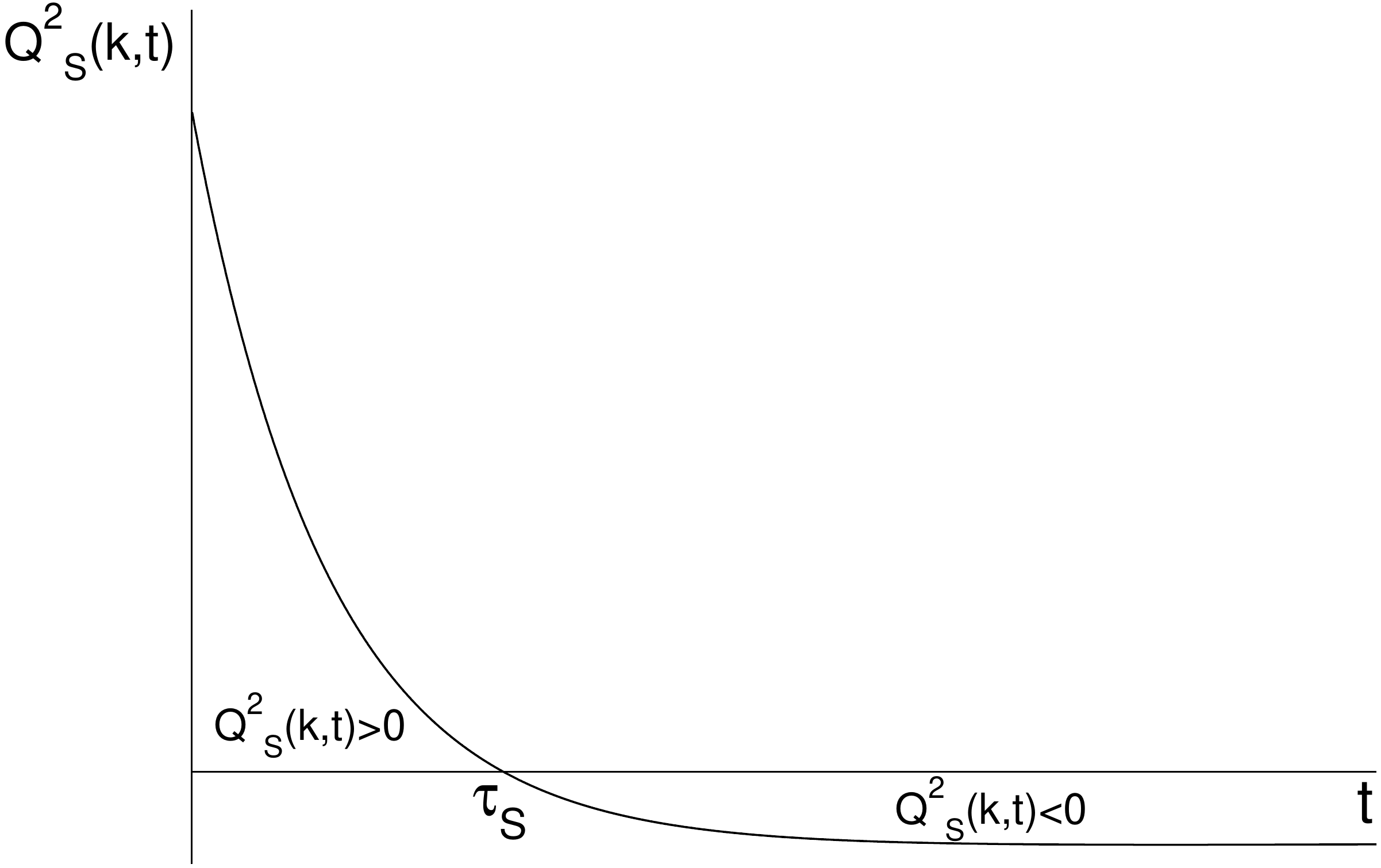}}
\subfigure[]{
\label{ch2:QSb}
\includegraphics[scale=0.35]{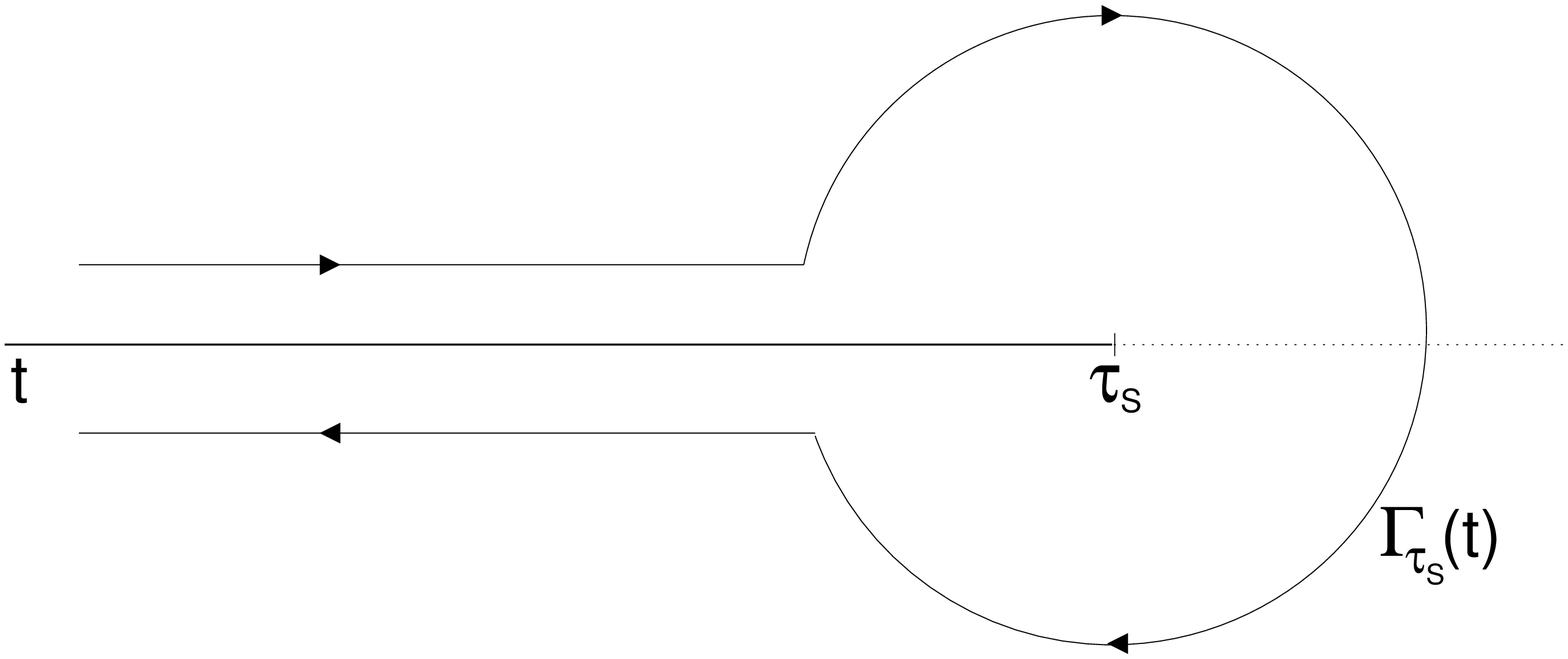}}
\subfigure[]{
\label{ch2:QSc}
\includegraphics[scale=0.35]{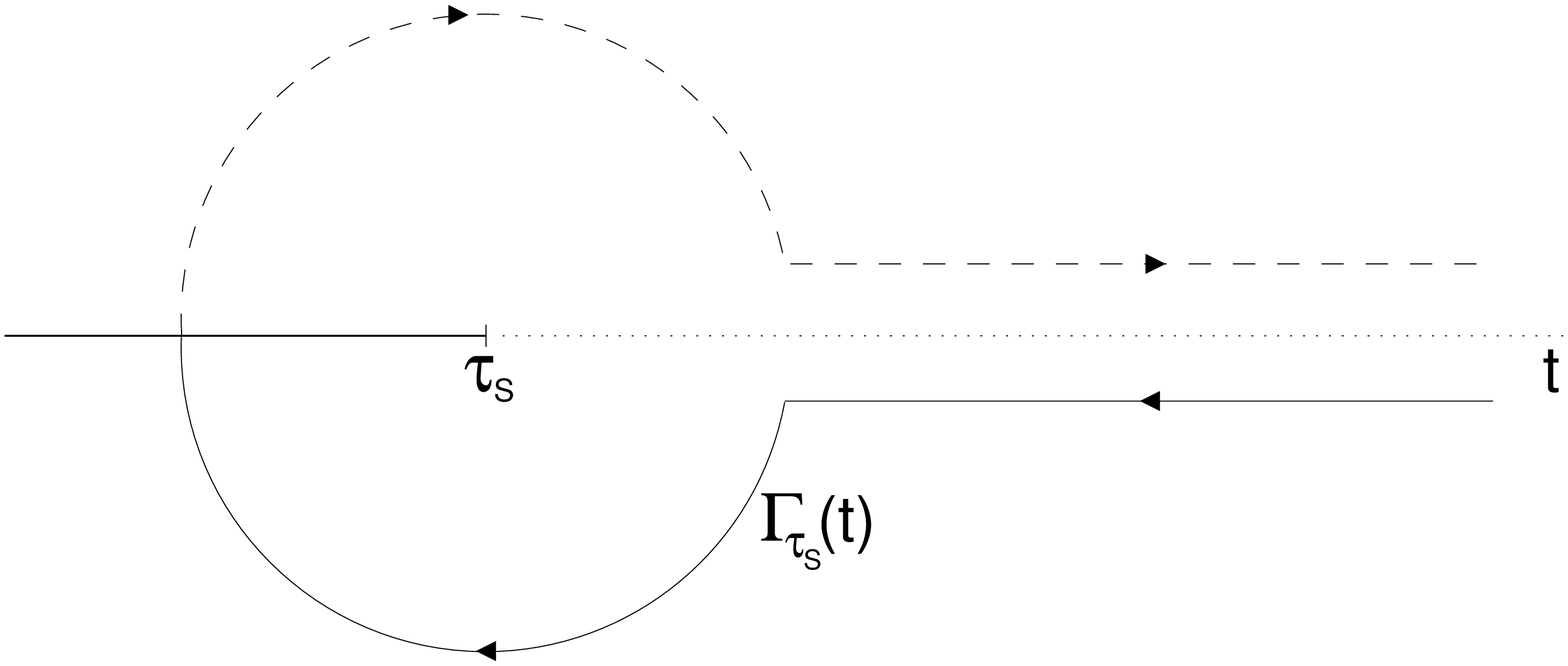}}
\caption{\small{(a) Behavior of the function  $Q_\sca^2(k,t)$. (b) Contour of integration $\Gamma_{\tau_\sca}(t)$ for  $0<t<\tau_\sca$. (c) Contour of  integration  $\Gamma_{\tau_\sca}(t)$ for $t>\tau_\sca$. The dashed line indicates the part of the path on the second Riemann sheet.}}
\end{center}
\end{figure}

\begin{figure}[htbp]
\begin{center}
\subfigure[]{
\label{ch2:QTa}
\includegraphics[scale=0.4]{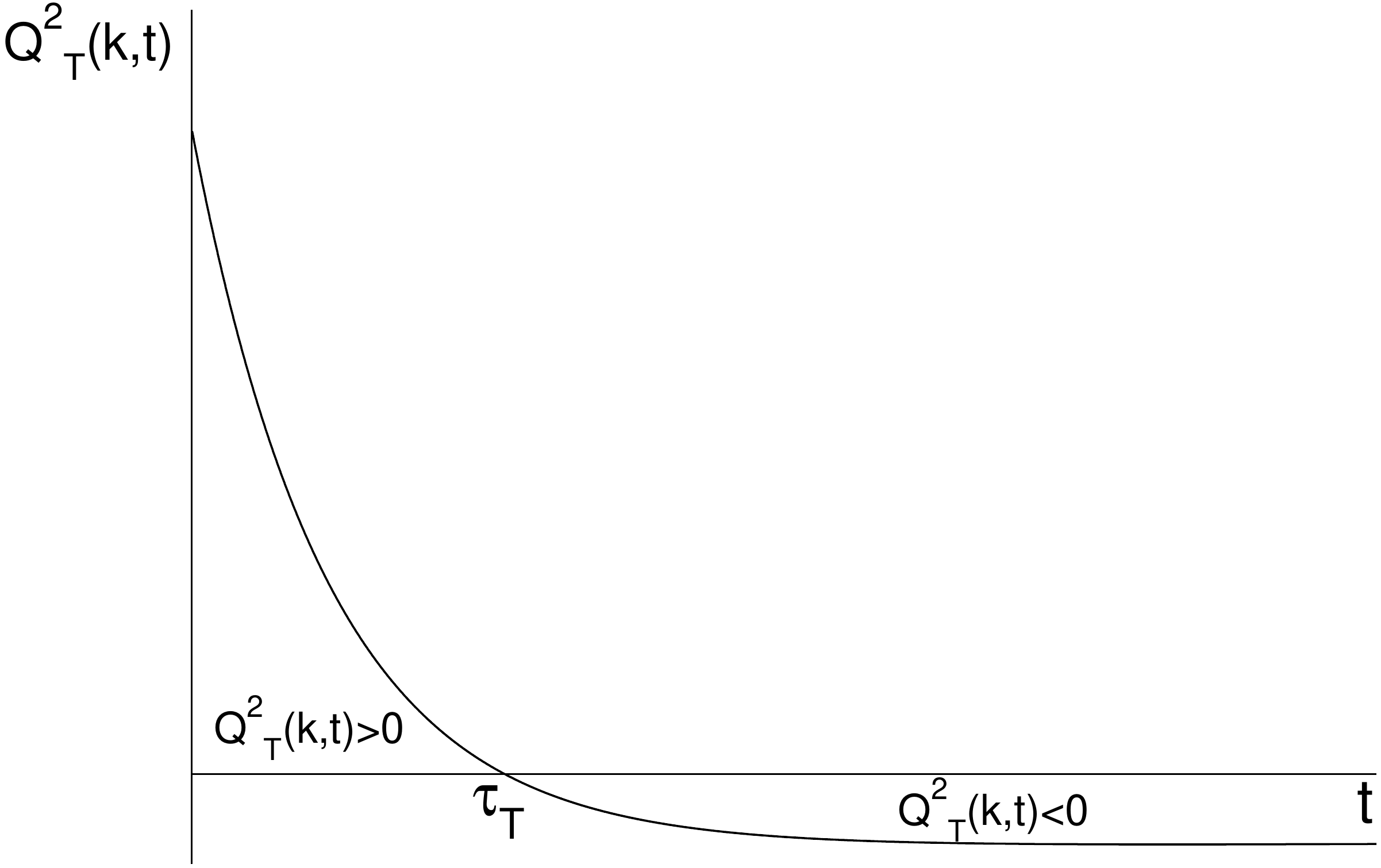}}
\subfigure[]{
\label{ch2:QTb}
\includegraphics[scale=0.35]{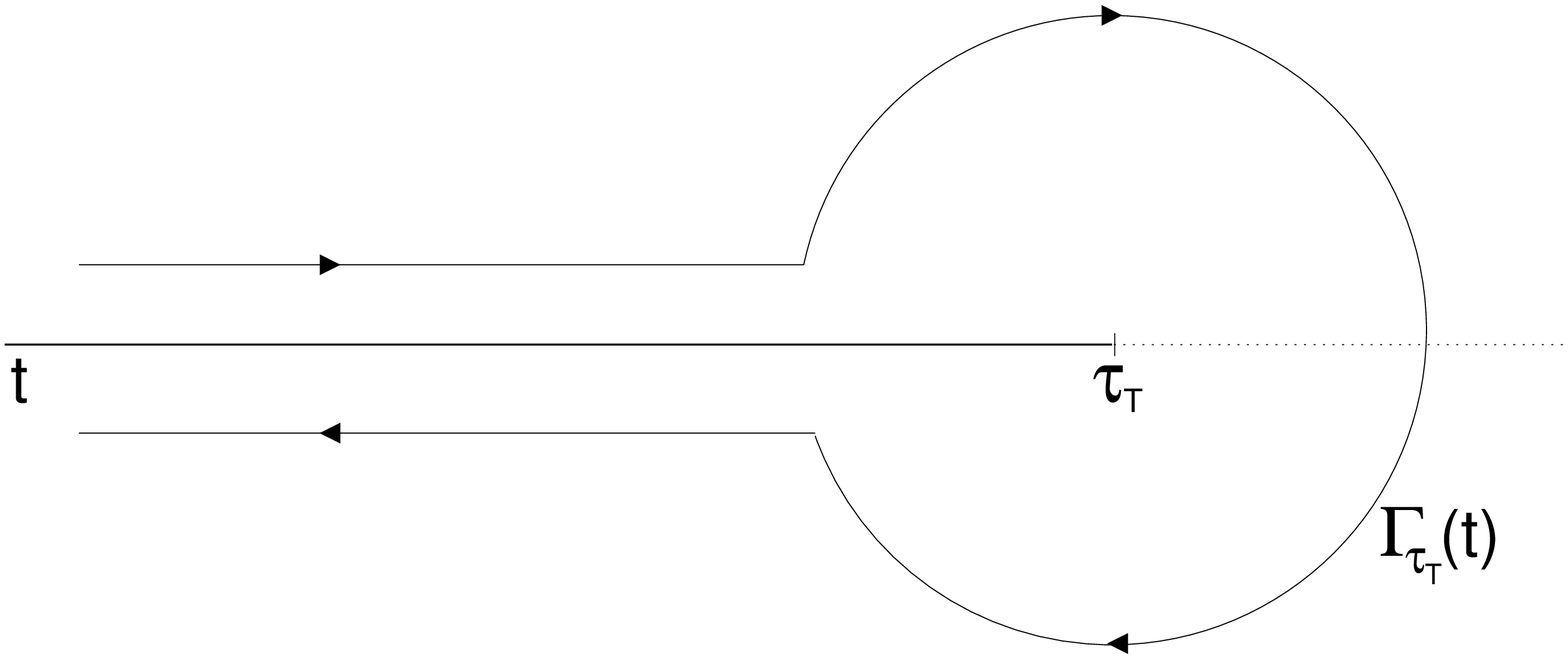}}
\subfigure[]{
\label{ch2:QTc}
\includegraphics[scale=0.35]{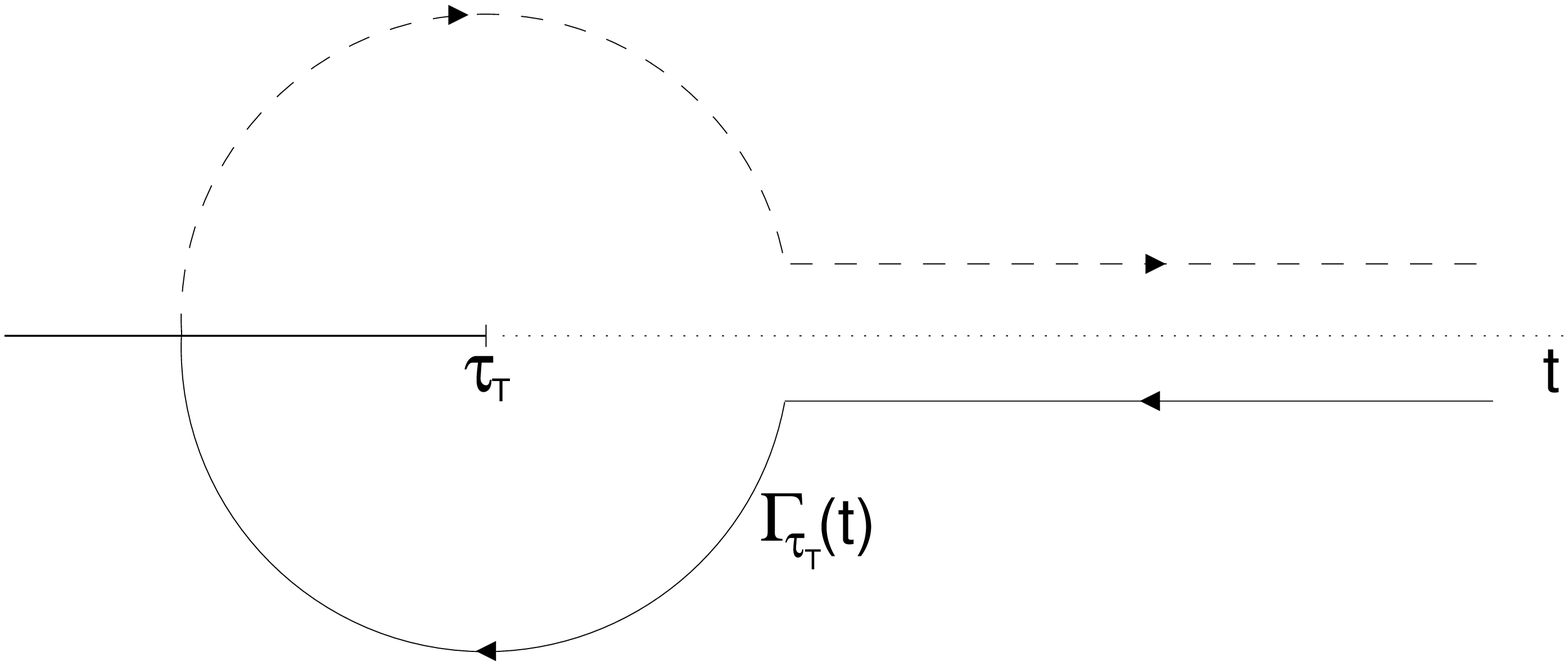}}
\caption{\small{(a) Behavior of  $Q_\ten^2(k,t)$.
(b) Contour of integration $\Gamma_{\tau_\ten}(t)$ for $0<t<\tau_\ten$. 
(c) Contour of integration  $\Gamma_{\tau_\ten}(t)$ for $t>\tau_\ten$. 
The dashed lined indicates the part of the path on the second Riemann sheet.}}
\end{center}
\end{figure}

In order to solve Eq. \eqref{ch2_ddotUk} and  Eq. \eqref{ch2_ddotVk}   with the help of the phase-integral   method, we choose the following base functions $Q_{\sca,\ten}$ for the scalar and tensor perturbations

\begin{eqnarray}
\label{Q}
Q_\sca^2(k,t)&=&R_\sca(k,t),\\
Q_\ten^2(k,t)&=&R_\ten(k,t),
\end{eqnarray}
where  $R_\sca(k,t)$ and  $R_\ten(k,t)$ are given by  Eq. \eqref{RS} and \eqref{RT}, respectively.  Using this selection, the phase-integral   method is  valid as  $k t\rightarrow \infty$, limit where we should impose the condition \eqref{ch2_cero_Uk}, where the validity condition   $\mu \ll 1$ holds. The selection, given in Eq. (\ref{Q}), makes the first order phase-integral  method coincide with the WKB solution. The bases functions  $Q_\sca(k,t)$ and  $Q_\ten(k,t)$ possess turning points  $t_\ret=\tau_\sca=1.33414\times 10^6 \,M_\Pl^{-1}$ for the mode $k=0.05\,\Mpc^{-1}$ and $t_\ret=\tau_\ten=833107 \,M_\Pl^{-1}$ for the mode $k=0.002\,\Mpc^{-1}$. The turning point represents the horizon.  There are two ranges  where to define the solution. To the left of the turning point  $0<t<t_\ret$ we have the classically permitted region $Q_{\sca,\ten}^2(k,t)>0$ and to the right of the turning point $t>t_\ret$ corresponding to the classically forbidden region
$Q_{\sca,\ten}^2(k,t)<0$, such as it is shown in Figs \ref{ch2:QSa}  and Fig. \ref{ch2:QTa}.

The mode  $k$ equations for the scalar an tensor perturbations (\ref{ch2_ddotUk})  and  (\ref{ch2_ddotVk}) in the phase-integral   method has two solutions:  For $0<t <t_\ret$

\begin{eqnarray}
\label{ch2_uk_left}
u^\phai_k(t)&=& \frac{c_1}{\sqrt{a(t)}}\left|q_\sca^{-1/2}(k,t)\right| \cos{\left[\left|\omega_\sca(k,t)\right|-\frac{\pi}{4}\right]} \\
\nonumber
&+& \frac{c_2}{\sqrt{a(t)}}\left|q_\sca^{-1/2}(k,t)\right| \cos{\left[\left|\omega_\sca(k,t)\right|+\frac{\pi}{4}\right]},\\
\label{ch2_vk_left}
v^\phai_k(t)&=& \frac{d_1}{\sqrt{a(t)}}\left|q_\ten^{-1/2}(k,t)\right| \cos{\left[\left|\omega_\ten(k,t)\right|-\frac{\pi}{4}\right]} \\
\nonumber
&+ &\frac{d_2}{\sqrt{a(t)}} \left|q_\ten^{-1/2}(k,t)\right| \cos{\left[\left|\omega_\ten(zk,t\right|+\frac{\pi}{4}\right]}.
\end{eqnarray}

and for $t>t_\ret$

\begin{eqnarray}
\label{ch2_uk_right}
u^\phai_k(t)&=&\frac{c_1}{2\sqrt{a(t)}}\left|q_\sca^{-1/2}(k,t)\right|\exp\left[-\left|\omega_\sca(k,t)\right|\right]\\
\nonumber
&+& \frac{c_2}{\sqrt{a(t)}} \left|q_\sca^{-1/2}(k,t)\right| \exp\left[\left|\omega_\sca(k,t)\right|\right],\\
\label{ch2_vk_right}
v^\phai_k(z)&=&\frac{d_1}{2\sqrt{a(t)}}\left|q_\ten^{-1/2}(k,t)\right|\exp\left[-\left|\omega_\ten(k,t)\right|\right]\\
\nonumber
& +& \frac{d_2}{\sqrt{a(t)}} \left|q_\ten^{-1/2}(k,t)\right| \exp\left[\left|\omega_\ten(k,t)\right|\right].
\end{eqnarray}
Notice that  Eq.  \eqref{limit_uk} and Eq. \eqref{limit_vk} are identical to Eq.  \eqref{ch2_uk_right} and Eq. \eqref{ch2_vk_right}  obtained in the first-order phase-integral   method.

Using the phase-integral  method up to third order ($2 n+1=3\rightarrow n=1$), we have that $q_{\sca}(k,t)$ and  $q_{\ten}(k,t)$ can be expanded in the form

\begin{eqnarray}
\label{q1}
q_\sca(k,t)=\sum_{n=0}^1 Y_{2n_\sca}(k,t) Q_\sca(k,t)=\left[Y_{0_\sca}(k,t)+Y_{2_\sca}(k,t)\right] Q_\sca(k,t),\\
\label{q2}
q_\ten(k,t)=\sum_{n=0}^1 Y_{2n_\ten}(k,t)Q_\ten(k,t)=\left[Y_{0_\ten}(k,t)+Y_{2_\ten}(k,t) \right] Q_\ten(k,t).
\end{eqnarray}

In order to compute $q_{\sca,\ten}(k,t)$, we  compute $Y_{2_{\sca,\ten}}(k,t)$ and   the required function  $\varepsilon_{0_{\sca,\ten}}(k,t)$.  The expressions  (\ref{q1}) and \eqref{q2} give  a third-order approximation for $q_{\sca,\ten}(k,t)$.  In order to compute $\omega_{\sca,\ten}(k,t)$  we make a contour integration following the path indicated in Fig. \ref{ch2:QTb}-(c).

\begin{eqnarray}
\nonumber
\omega_\sca(k,t)&=&\omega_{0_\sca}(k,t)+ \omega_{2_\sca}(k,t),\\
\nonumber
&=&\int_{\tau_\sca}^{t}Q_\sca(k,t)\D t+\frac{1}{2}\int_{\Gamma_{\tau_\sca}}Y_{2_\sca}(k,t)Q_\sca(k,t)\D t,\\
&=&\int_{\tau_\sca}^{t}Q_\sca(k,t)\D t+\frac{1}{2}\int_{\Gamma_{\tau_\sca}}f_{2_\sca}(k,t)\D t.
\end{eqnarray}

\begin{eqnarray}
\nonumber
\omega_\ten(k,t)&=&\omega_{0_\ten}(k,t)+ \omega_{2_\ten}(k,t),\\
\nonumber
&=&\int_{\tau_\ten}^{t}Q_\ten(k,t)\D t+\frac{1}{2}\int_{\Gamma_{\tau_\ten}}Y_{2_\ten}(k,t)Q_\ten(k,t)\D t,\\
&=&\int_{\tau_\sca}^{t}Q_\ten(k,t)\D z+\frac{1}{2}\int_{\Gamma_{\tau_\ten}}f_{2_\ten}(k,t)\D t,
\end{eqnarray}

where
\begin{eqnarray}
f_{2_\sca}(k,t)&=&Y_{2_\sca}(k,t)Q_\sca(k,t),\\
f_{2_\ten}(k,t)&=&Y_{2_\ten}(k,t)Q_\ten(k,t).
\end{eqnarray}
The functions  $f_{2_\sca}(k,t)$ and  $f_{2n_\ten}(k,t)$ have the following functional dependence:

\vspace{-0.5cm}
\begin{eqnarray}
\label{A}
f_{2_\sca}(k,t)&=&A(k,t)(t-\tau_\sca)^{-5/2},\\
\label{B}
f_{2_\ten}(k,t)&=&B(k,t)(t-\tau_\ten)^{-5/2},
\end{eqnarray}
where the functions $A(k,t)$ is regular at $\tau_\sca$, and the function  $B(k,t)$   is regular at  $\tau_\ten$. With the help of  the functions  \eqref{A}-\eqref{B} we compute the integrals for $\omega_{2_{\sca,\ten}}$ using the contour indicated in  Figs. \ref{ch2:QSb}-(c) and \ref{ch2:QTb}-(c). The expressions for  $\omega_{2_{\sca,\ten}}$ permit one to obtain the third-order phase integral approximation of the solution to the equations for scalar \eqref{ch2_ddotUk} and tensor \eqref{ch2_ddotVk} perturbations.  The constants $c_1$, $c_2$, $d_1$ and  $d_2$ are obtained using the limit  $k\,t\rightarrow 0$ of the solutions on the left side of the turning point \eqref{ch2_uk_left} and \eqref{ch2_vk_left}, and  are given by the expressions

\begin{eqnarray}
c_1&=&-\im\,c_2,\\
c_2&=&\frac{\e^{-\im\frac{\pi}{4}}}{\sqrt{2}}\e^{-\im\left[k\,\eta(0)+\left|\omega_{0_\sca}(k,0)\right|\right]},\\
d_1&=&-\im\,d_2,\\
d_2&=&\frac{\e^{-\im\frac{\pi}{4}}}{\sqrt{2}}\e^{-\im\left[k\,\eta(0)+\left|\omega_{0_\ten}(k,0)\right|\right]},
\end{eqnarray}
In order to compute the scalar and tensor power spectra, we need to calculate the limit as  $k\,t\rightarrow \infty$ of the growing part of the solutions on the right side of the turning point  given by   Eq. \eqref{ch2_uk_right} and  Eq. \eqref{ch2_vk_right} for scalar and tensor perturbations respectively.

\begin{eqnarray}
\label{PS_pi}
P_\sca(k)&=&\lim_{-k t\rightarrow \infty} \frac{k^3}{2\pi^2} \left|\frac{u_k^\phai(t)}{z_\sca(t)}\right|^2,\\
\label{PT_pi}
P_\ten(k)&=&\lim_{-k t\rightarrow \infty} \frac{k^3}{2\pi^2} \left|\frac{v_k^\phai(t)}{a(t)}\right|^2.
\end{eqnarray}

\subsection{Numerical Integration}

The equation for scalar and tensor perturbations \eqref{dotu} and \eqref{dotv} are integrated numerically, they are set using the expressions for  $a_\ex(t)$ and $\phi_\ex(t)$. The perturbations $u_k$ and $v_k$ are complex functions, then two differential equations are solved for each one, the equation for the real part and the equation for the imaginary part. 

The integration is done in two parts: the first part is done in the limit when $k^2\gg\sfrac{\left(\dot{a}\dot{z_\sca}+a\ddot{z_\sca}\right)a}{z_\sca} $, and $k^2\gg \dot{a}^2+a\ddot{a}$ for scalar and tensor perturbations, respectively. In the second part, the full equations \eqref{dotu}  and \eqref{dotv} are considered. The first part corresponds with the time when perturbations are inside the horizon, then$u_k$ and  $v_k$ exhibits an oscillatory behavior,

\begin{eqnarray}
\label{dotuk_k2}
\ddot{u_k}&+\dfrac{\dot{a}}{a}\dot{u_k}+\dfrac{k^2}{a^2} u_k=0,\\
\label{dotvk_k2}
\ddot{v_k}&+\dfrac{\dot{a}}{a}\dot{v_k}+\dfrac{k^2}{a^2} v_k=0,
\end{eqnarray}
from 300 to 100 oscillations before the horizon crossing, using as initial condition equation \eqref{boundary_0}. Then, we use the final stage of this solution as initial condition, to solve equations \eqref{dotu} and \eqref{dotv}  from 100 oscillations before horizon crossing to roughly three times the horizon crossing time when the perturbation is frozen.  
Finally, with Eqs. \eqref{PS} and \eqref{PT}  we calculate numerically the scalar and tensor perturbation. 

\begin{figure}[htp]
\subfigure[]{\label{fig:a}\includegraphics[scale=0.27]{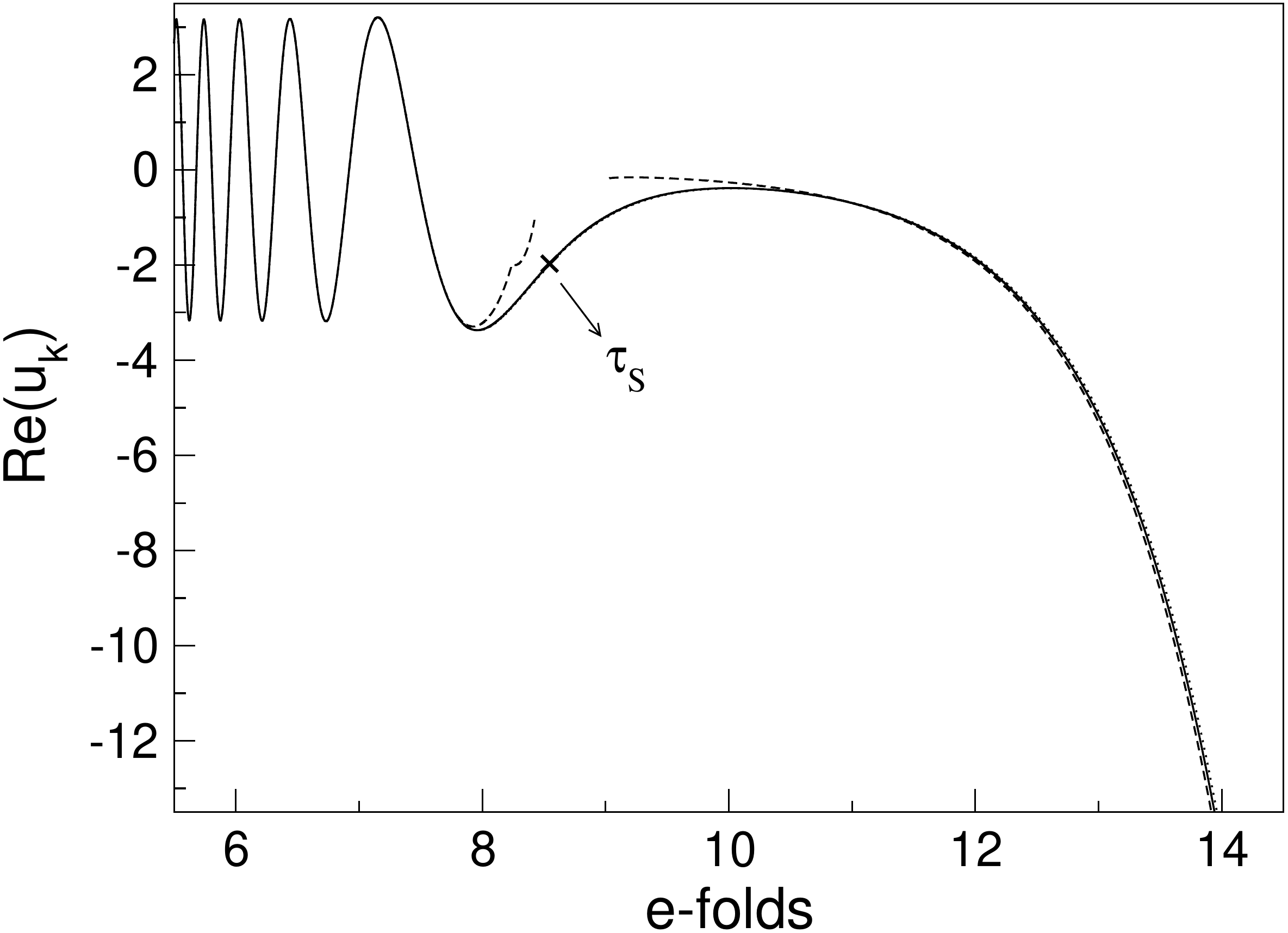}}
\subfigure[]{\label{fig:b}\includegraphics[scale=0.27]{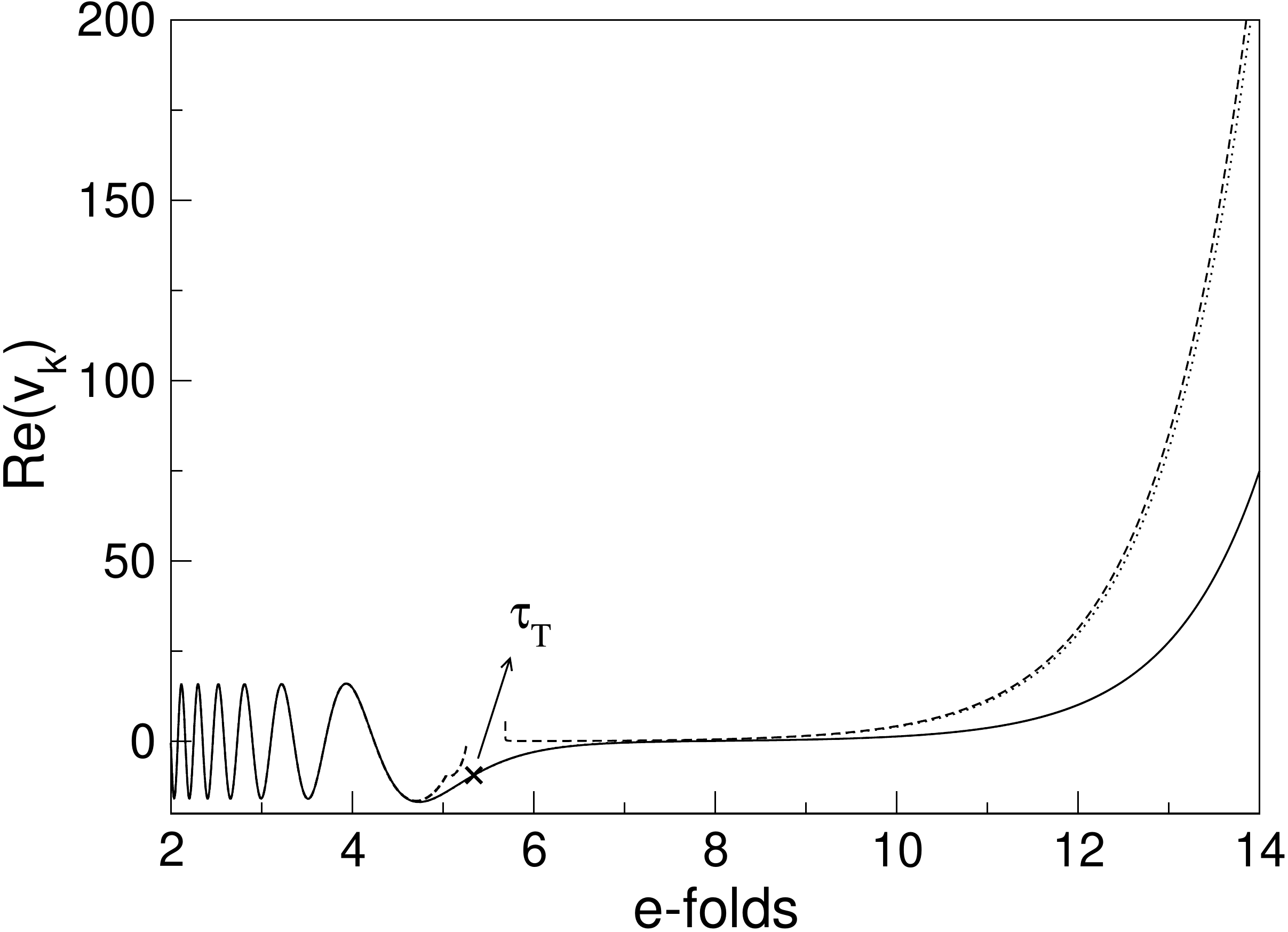}}
\caption{(a) $\textnormal{Re}\left(u_k\right)$  for $k=0.05$ and,  (b) $\textnormal{Re}\left(v_k\right)$ for $k=0.002$  versus the number of e-folds for the generalized Starobinsky inflationary model, where  solid line represents the numerical solution,  dashed line the third-order phase-integral   method, and dotted-line the second-order uniform approximation method.}
\label{Re}
\end{figure}

\begin{figure}[htp]
\subfigure[]{\label{fig:a}\includegraphics[scale=0.27]{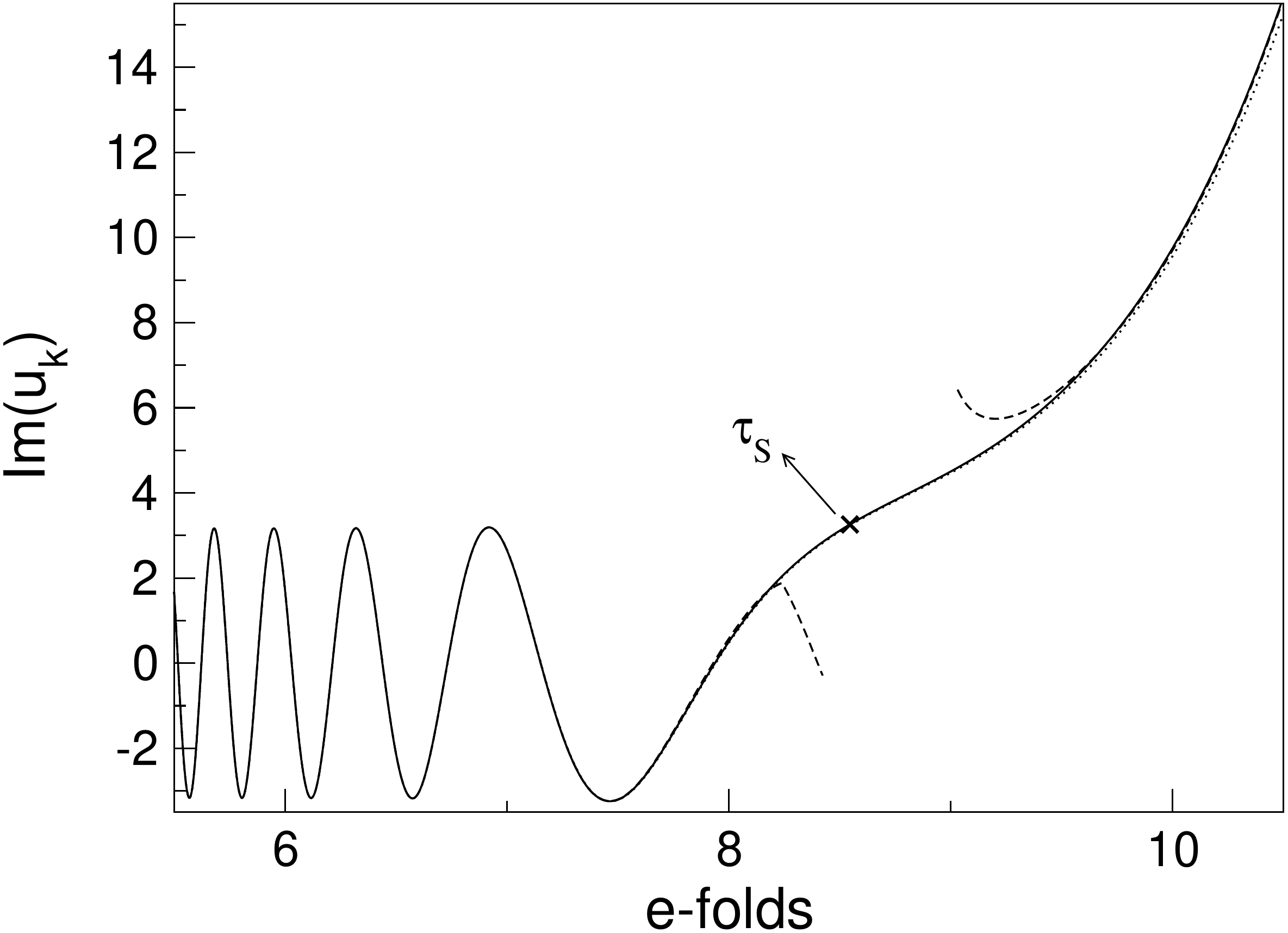}}
\subfigure[]{\label{fig:b}\includegraphics[scale=0.27]{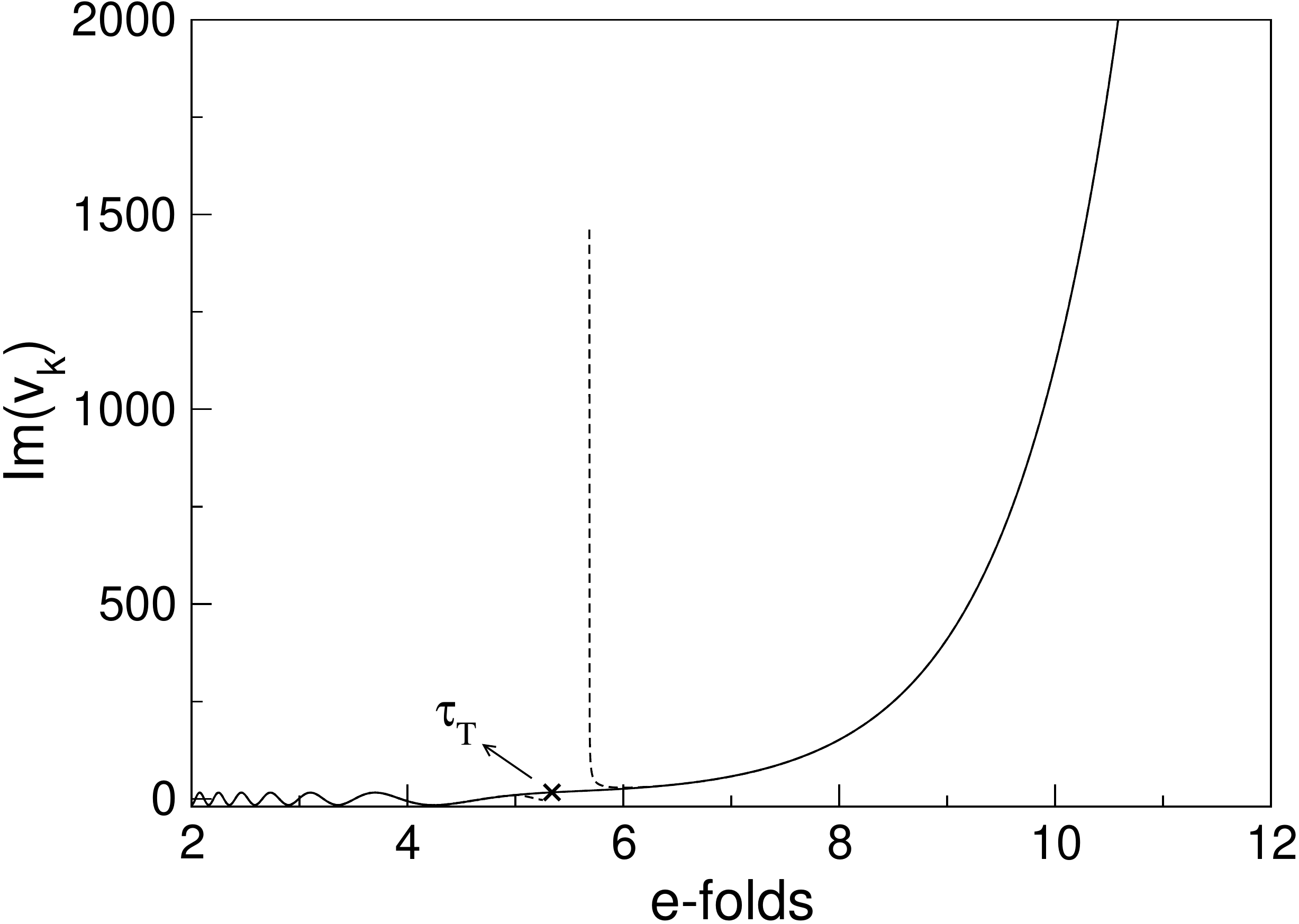}}
\caption{(a) $\textnormal{Im}\left(u_k\right)$ for $k=0.05$ and,  (b) $\textnormal{Im}\left(v_k\right)$ for $k=0.002$ versus the number of e-folds for the generalized Starobinsky inflationary model, where  solid line represents the numerical solution,  dashed line the third-order phase-integral approximation method, and dotted-line the second-order uniform approximation method.}
\label{Im}
\end{figure}

\begin{figure}[htp]
\subfigure[]{\label{fig:a}\includegraphics[scale=0.27]{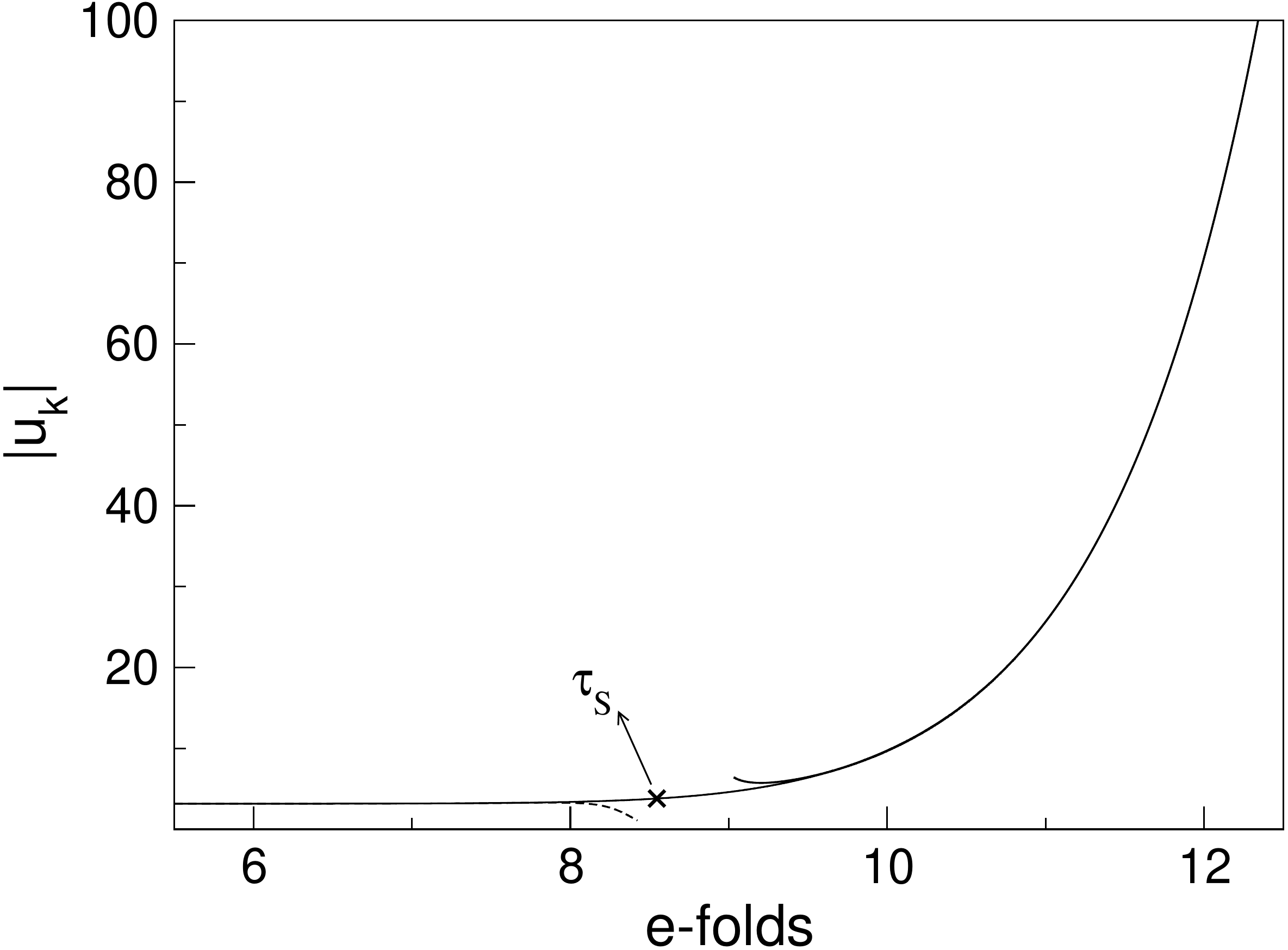}}
\subfigure[]{\label{fig:b}\includegraphics[scale=0.27]{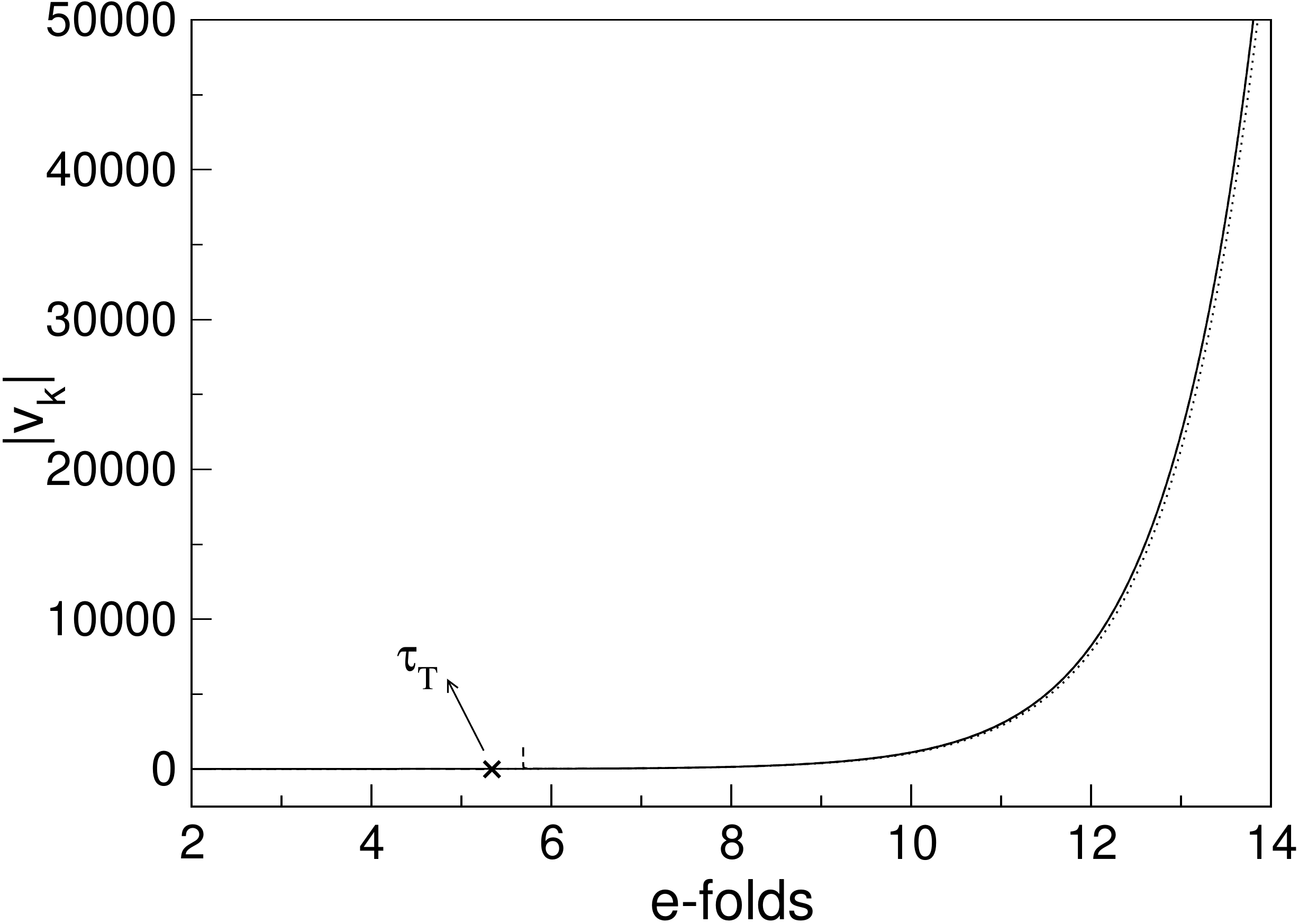}}
\caption{(a) $\textnormal{Abs}\left(u_k\right)$ for $k=0.05$, and  (b) $\textnormal{Abs}\left(v_k\right)$ for $k=0.002$  versus the number of e-folds for the generalized Starobinsky inflationary model, where  solid line represents the numerical solution, dashed line the third-order phase-integral approximation method, and dotted-line the second-order uniform approximation method.}
\label{Abs}
\end{figure}

Figs. \ref{Re}(a) and \ref{Re}(b) show the real part of the scalar a tensor perturbations calculated numerically and with the semiclassical methods described before. Figs. \ref{Im}(a) and \ref{Im}(b) show the imaginary part of the scalar and tensor perturbations. Finally, in Fig. \ref{Abs}(a) and \ref{Abs}(b) we can observed the behaviour of absolute value of the perturbations. Note that the real part of the tensor perturbations calculated with semiclassical methods moves away the numerical result for a number of e-folding $N>8$.

\section{Results and Discussion}

\begin{figure}[th!]
\subfigure[]{\label{fig:a}\includegraphics[scale=0.25]{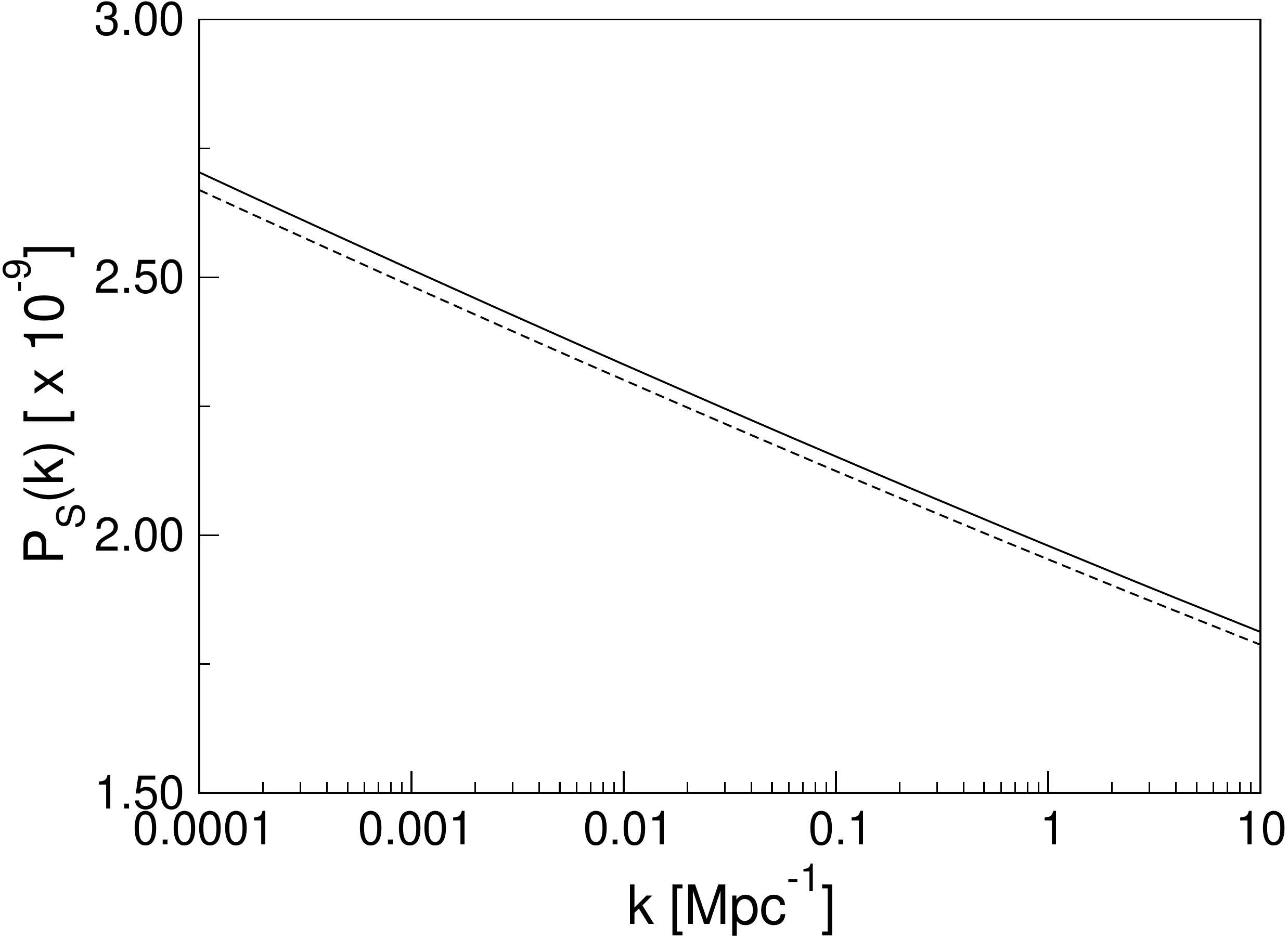}}
\subfigure[]{\label{fig:b}\includegraphics[scale=0.25]{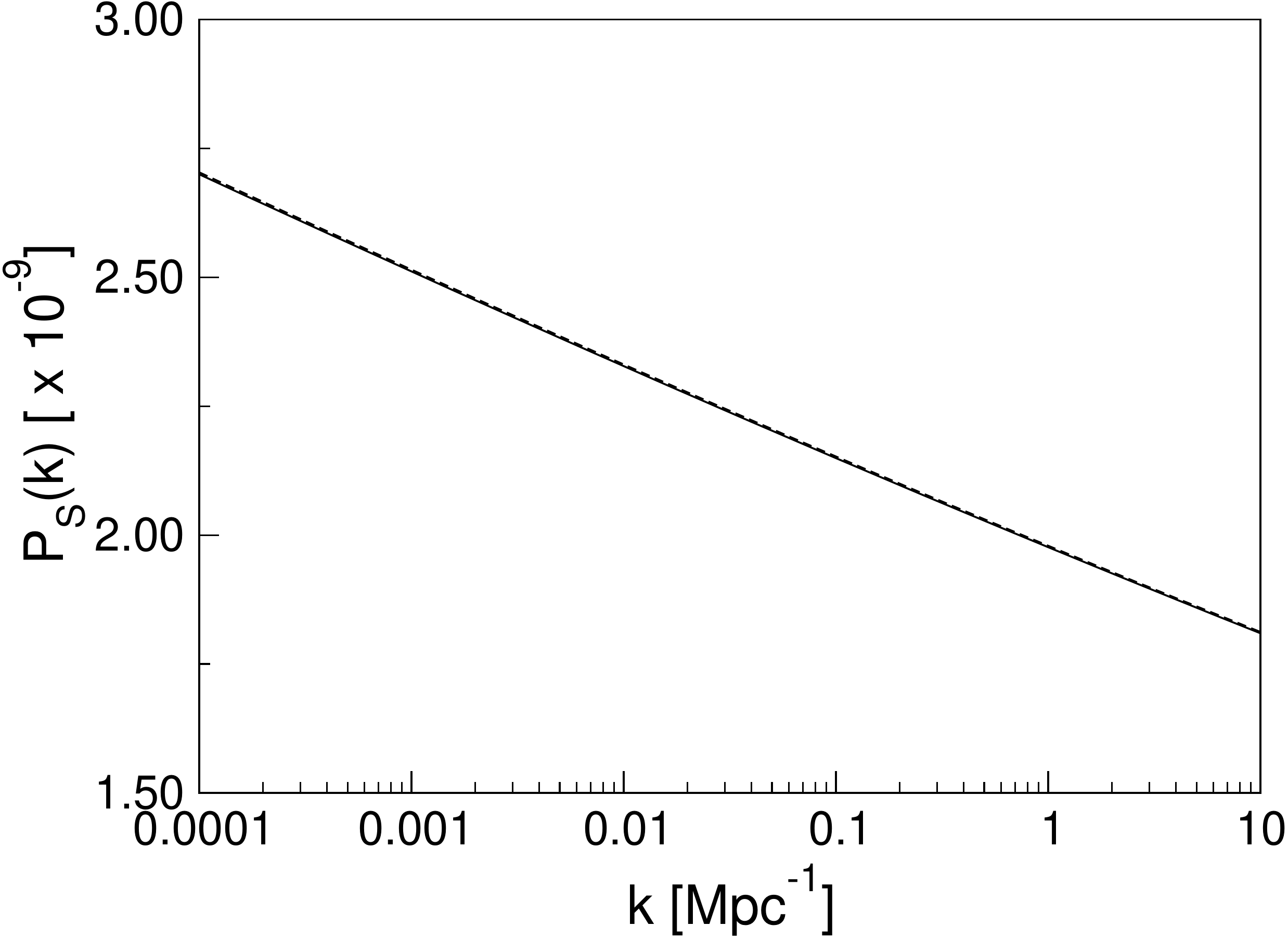}}
\caption{Evolution of $P_\sca(k)$  for the generalized Starobinsky inflationary model respect to $k$, where  solid line represents numerical solution and  dashed line the approximation methods: a) Second-order uniform approximation method, and b) Third-order phase-integral  method.}
\label{PSpi3_graph}
\end{figure}

\begin{figure}[th!]
\subfigure[]{\label{fig:a}\includegraphics[scale=0.25]{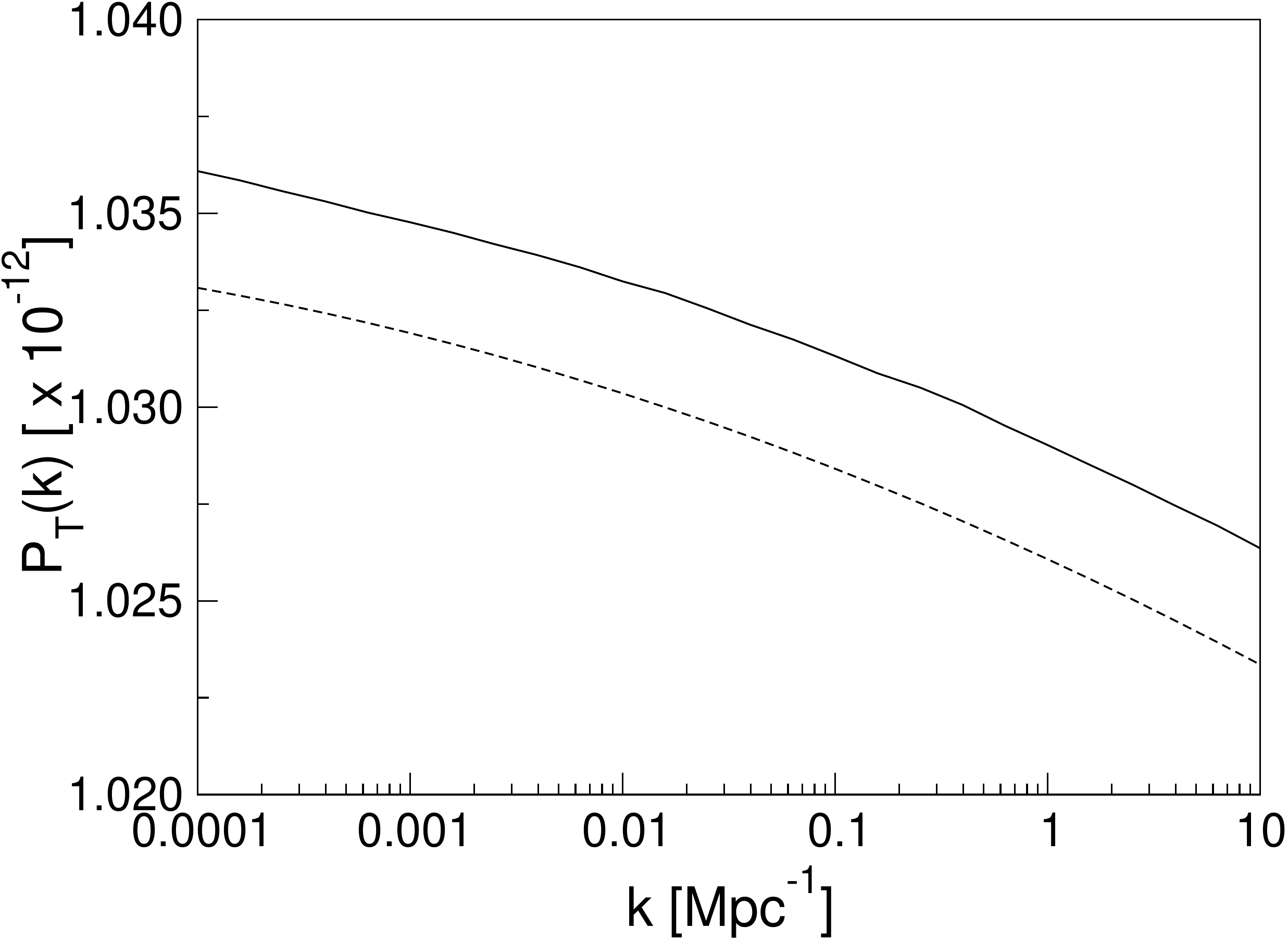}}
\subfigure[]{\label{fig:b}\includegraphics[scale=0.25]{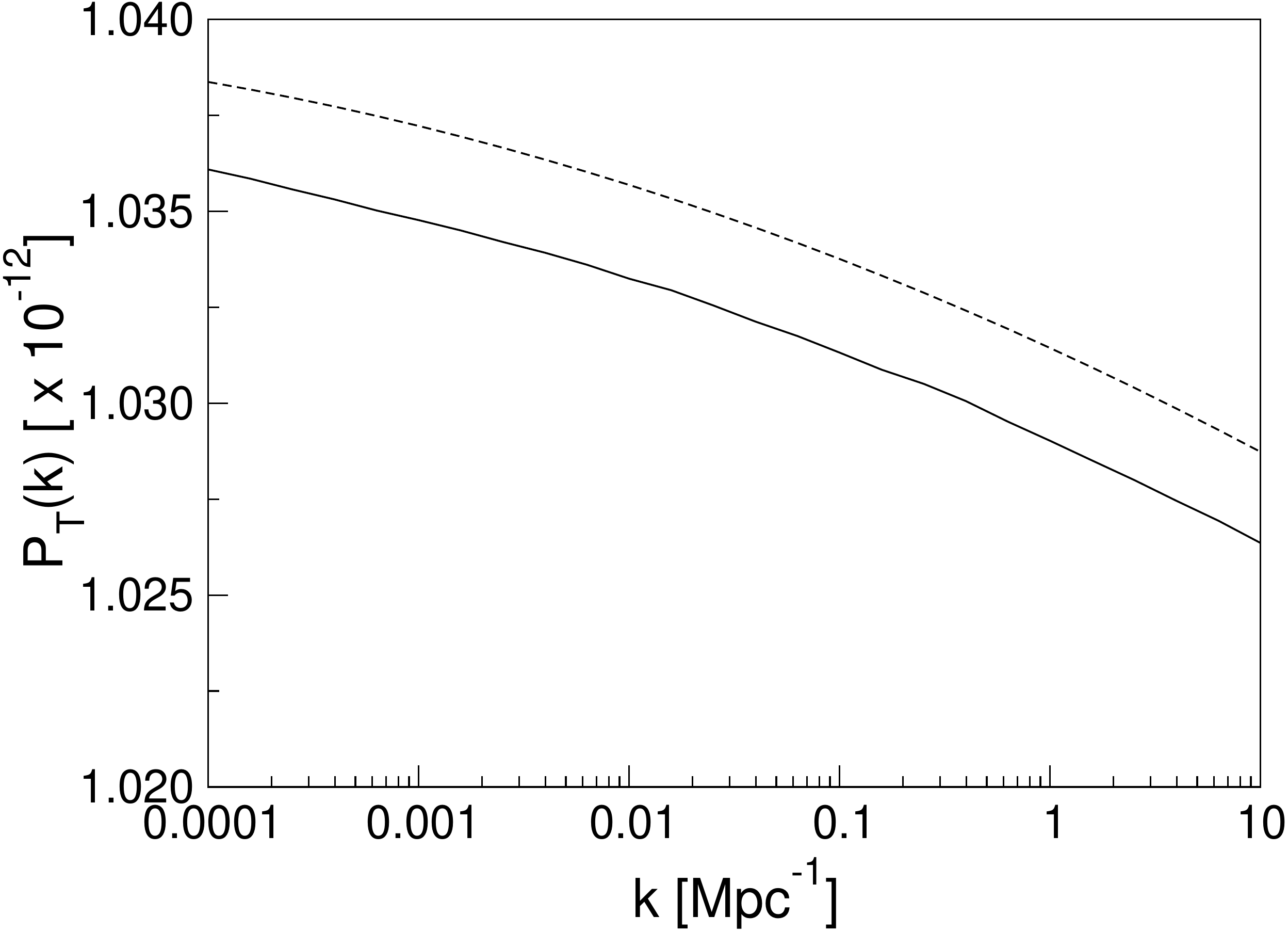}}
\caption{Evolution of $P_\ten(k)$  for the generalized Starobinsky inflationary model respect to $k$, where  solid line represents numerical solution and  dashed line the approximation methods:  a) Second-order uniform approximation method, and b) Third-order phase-integral  method.}
\label{PTpi3_graph}
\end{figure}

Using semiclassical methods we have obtained the dependence of the scalar power spectrum $P_\sca$  and the tensor power spectrum $P_\ten$ in terms of $k$ in  the range $0.0001\, \textnormal{Mpc}^{-1}  \leq k \leq 10\, \textnormal{Mpc}^{-1}$.  These results are compared with the numerical solution, which are shown in Fig. \ref{PSpi3_graph} and Fig. \ref{PTpi3_graph}.

The scalar power spectrum  $P_\sca$ satisfies the following power-law \cite{akrami:2018b}:

\begin{equation}
\ln P_\sca=\ln A_\sca + \left(n_S -1\right) \ln\left(\dfrac{k}{k_*} \right),
\end{equation}
where $A_\sca$ is the scalar power spectrum amplitude, and $k_{*}$ is the pivot scale. Because $n_S$ is close to the unity, in the following we are going to report $A_\sca$ instead $P_\sca$.
Using the slow-roll approximation and the semiclassical methods we calculate the cosmological parameters: $A_\sca$, $n_\sca$, and $r$.
Table \ref{p_1.0004_table_sr} shows the comparation of the cosmological parameters calculated numerically and with the  second-order slow-roll approximation, whereas Tables \ref{p_1.0004_table_ua2} and \ref{p_1.0004_table_pi3} show the comparation between the numerical  solution with those  obtained using the second-order uniform approximation method and with the phase-integral approximation  method up-to third-order in deviation, respectively. The values of  $A_\sca$ and $n_\sca$ are calculated at the pivot scale $k=0.05$ Mpc$^{-1}$.

\begin{table}[htbp]
\begin{center}
\begin{tabular}{cccc}
\toprule
Parameter                                           & Numerical & $2^\textnormal{nd}$-order  $sr$ approximation&rel. err (\%) \\  
\midrule 
$A_\sca$                                           & $\;\;2.2024 \times 10^{-9}$           &   $\;\;2.1576 \times 10^{-9}$& $2.0378$\\
$\ln\left(10^{10} A_\sca \right) $   & $\;\;3.092$   & $\;\;3.071$& $0.665$\\
$n_\sca$                                            & $\;\;0.9655$ &$\;\;0.9627$&  $0.2927$\\
$r_{0.002} $                                     & $\;\;0.00338$ & $\;\;0.00350$&$3.62673$ \\
\bottomrule
\end{tabular}
\caption{Cosmological parameters for the  generalized Starobinsky inflationary model for $p=1.0004$ calculated: $a)$ numerically and $b)$  the second-order  slow-roll approximation. Here $sr$ means slow-roll.}
\label{p_1.0004_table_sr}
\end{center}
\end{table}

\begin{table}[th!]
\begin{center}
\begin{tabular}{cccc}
\toprule
Parameter                                           & Numerical &$2^\textnormal{nd}$-order  $ua$ method&rel. err (\%) \\  
\midrule 
$A_\sca$                                           & $\;\;2.2024 \times 10^{-9}$           &   $\;\;2.1768 \times 10^{-9}$ & $\;\;1.1645$   \\
$\ln\left(10^{10} A_\sca \right) $   & $\;\;3.092$   &$\;\;3.080$& $\;\;0.379$\\
$n_\sca$                                            & $\;\;0.9655$ &$\;\;0.9655$ &$ \;\;0.0083$ \\
$r_{0.002} $                                     & $\;\;0.00338$ & $\;\;0.00341$& $\;\;0.80727$\\
\bottomrule
\end{tabular}
\caption{Cosmological parameters for the  generalized Starobinsky inflationary model for $p=1.0004$ calculated: $a)$ numerically and $b)$ the uniform approximation method up to second-order in deviation. Here $ua$ means uniform approximation.}
\label{p_1.0004_table_ua2}
\end{center}
\end{table}

\begin{table}[th!]
\begin{center}
\begin{tabular}{cccc}
\toprule
Parameter                                           & Numerical & $3^\textnormal{rd}$-order $pi$ method&rel. err (\%) \\  
\midrule 
$A_\sca$                                           & $\;\;2.2024 \times 10^{-9}$           &   $\;\;2.2056 \times 10^{-9}$ & $\;\;0.1436$   \\
$\ln\left(10^{10} A_\sca \right) $   & $\;\;3.092$   &$\;\;3.093$& $\;\;0.046$\\
$n_\sca$                                            & $\;\;0.9655$ &$\;\;0.9655$ &$ \;\;0.0002$ \\
			$r_{0.002} $                                     & $\;\;0.00338$ & $\;\;0.00338$& $\;\;0.01283$\\
\bottomrule
\end{tabular}
\caption{Cosmological parameters for the  generalized Starobinsky inflationary model for $p=1.0004$ calculated: $a)$ numerically and $b)$  the phase-integral method up to third-order in deviation. Here $pi$ means phase-integral.}
\label{p_1.0004_table_pi3}
\end{center}
\end{table}

From tables \ref{p_1.0004_table_sr}, \ref{p_1.0004_table_ua2}, and \ref{p_1.0004_table_pi3}, we can observed that for the three cosmological parameters  the phase-integral method up to third-order in deviation gives the smallest relative error.

\begin{figure}[th!]
\centering
\includegraphics[scale=0.7]{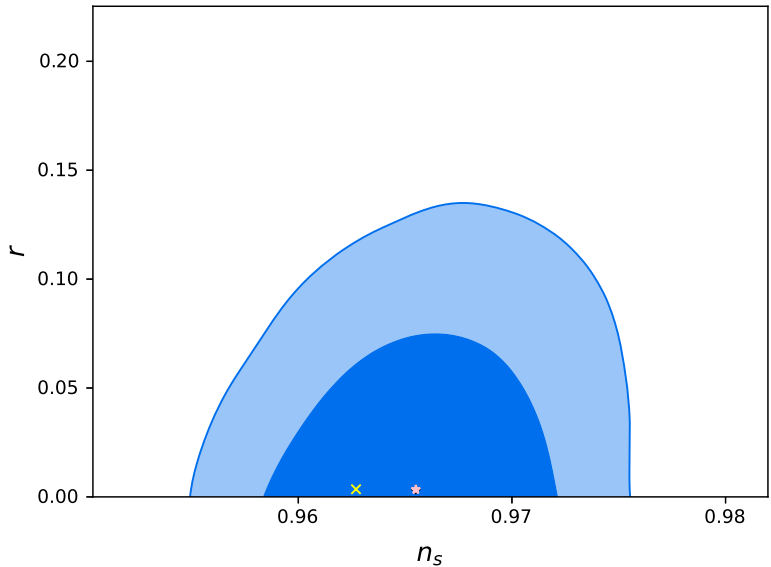}
\caption{Contour plot of $r$ vs $n_S$.  Here: yellow x represents the slow-roll approximation whereas the pink star semiclassical methods. }
\label{contour}
\end{figure}

In Fig. \ref{contour} we show the $(n_S,r)$ plane, where the blue contours correspond to the $68\%$ and $95\%$ CL results from Planck 2018 TT,TE,EE+lowE+lensing data \cite{akrami:2018a}. We can observe that the results obtained with semiclassical methods are inside the $95\%$ of confidence level.

\section{Conclusions}

 We calculated the scalar and tensor power spectra for the generalized Starobinsky inflationary model using semiclassical methods.We reported the behaviour of perturbations in terms of the number of e-folds. We found that the phase-integral method reproduces the scalar power spectrum $A_\sca$ with a relative error  of $0.1436\%$, and for the tensor-to-scalar ratio $r$ with a relative error of $0.01283\%$. Instead the uniform approximation method reproduces the scalar power spectrum $A_\sca$ with a relative error  of $1.1645\%$, and for the tensor-to-scalar ratio $r$ with a relative error of $0.80727\%$. In the  contour plot $r$  vs $n_S$ we can observe that our results are inside the $95\%$ of confidence level.

\section{Acknowledgment}
The author thanks to Dr. Werner B\"amer-Escamilla for doing the fitting of the scale factor $a$ and the scalar field $\phi$ using gnuplot \cite{gnuplot}.


\bibliographystyle{unsrt}

\begin{thebibliography}{10}
	
	\bibitem{guth:1981}
A.~H. Guth.
\newblock {Inflationary universe: A possible solution to the horizon and
	flatness problems}.
\newblock {\em Phys. Rev. D}, 23:347, 1981.

	\bibitem{tamayo:2017}
{D.A. Tamayo, J.A.S. Lima, M.E.S. Alves, J.C.N.de Araujo}
\newblock {Primordial gravitational waves in running vacuum cosmologies}.
\newblock {\em Astroparticle Physics}, 18-22:87, 2017.

\bibitem{tristram:2021}
{M. Tristram \textit{et al.}}
\newblock Planck constraints on the tensor-to-scalar ratio.
\newblock {\em Astronomy \& Astrophysics}, 647:A128.

\bibitem{ade:2021}
{P.A.R. Ade \textit{et al.}}
\newblock Improved constraints on primordial gravitational waves using planck,
wmap, and bicep/keck observations through the 2018 observing season.
\newblock {\em Phys. Rev. Lett.}, 127:151301.
8.

\bibitem{martin:2014}
{J. Martin, C. Ringeval, and V. Vennin}.
\newblock {Encyclopaedia Inflationaris}.
\newblock {\em Phys. Dark Univ.}, 5-6:75--235, 2014.

	
	\bibitem{starobinsky:1980}
	A.~A. Starobinsky.
	\newblock {A new type of isotropic cosmological models without singularity}.
	\newblock {\em Phys. Lett. B}, 91:99, 1980.
	
	\bibitem{truman:2021}
	{T. Tapia and C. Rojas}.
	\newblock {Semiclassical analysis of the tensor power spectrum in the
		Starobinsky inflationary model}.
	\newblock {\em Int. J. Mod. Phys. D}, 30:2150040, 2021.
	
	\bibitem{truman:2020}
	{T. Tapia, M. Z. Mughal, and C. Rojas}.
	\newblock {Semiclassical analysis of the Starobinsky inflationary model}.
	\newblock {\em Phys. Dark Univ.}, 30:100650, 2020.
	
	\bibitem{samart:2019}
	{D. Samart and P. Channuie}.
	\newblock {Unification of inflation and dark matter in the Higgs-Starobinsky
		model}.
	\newblock {\em Eur. Phys. J. C}, 79:347, 2019.
	
	\bibitem{adam:2019}
	{C. Adam and D. Varela}.
	\newblock {The superpotential method in cosmological inflation}.
	\newblock {\em arXiv:1901}, 2019.
	
	\bibitem{granada:2019}
	{L. N. Granada and D. F. Jimenez}.
	\newblock {Slow-roll inflation with exponential potential in scalar-tensor
		models}.
	\newblock {\em Eur. Phys. J. C}, 79:772, 2019.
	
	\bibitem{chowdhury:2019}
	{D. Chowdhury, J. Martin, C. Ringeval, and V. Vennin}.
	\newblock {Inflation after Planck: Judgment Day}.
	\newblock {\em arXiv:1902.03951}, 2019.
	
	\bibitem{paliathanasis:2017}
	A.~Paliathanasis.
	\newblock {Analytic solution of the Starobinsky model for inflation}.
	\newblock {\em Eur. Phys. J C}, 77:438, 2017.
	
	\bibitem{diValentino:2017}
	{E. Di Valentino and L. Mersini-Houghton}.
	\newblock {Testing predictions of the quantum landscape multiverse 1: the
		Starobinsky inflationary potential}.
	\newblock {\em JCAP}, 2, 2017.
	
	\bibitem{linde:2014}
	A.~Linde.
	\newblock {Inflationary Cosmology after Planck 2013}.
	\newblock {\em arXiv:1402.0526}, 2014.
	
	\bibitem{meza:2021}
	{S. Meza, D. Altamirano, M. Z. Mughal, and Clara Rojas}.
	\newblock {Numerical analysis of the generalized Starobinsky inflationary
		model}.
	\newblock {\em Int. J. Mod. Phys. D}, 30:2150062, 2021.
	
	\bibitem{renzi:2020}
	{F. Renzi, M. Shokri, and A. Melchiorri}.
	\newblock {What is the amplitude of the gravitational waves background expected
		in the Starobinsky model?}
	\newblock {\em Phys. Dark. Univ.}, 27:100450, 2020.
	
	\bibitem{canko:2020}
	{D. D. Canko, Ioannis D. Gialamas, and G. P. Kodaxis}.
	\newblock {A simple $F(\mathcal{R},\phi)$ deformation of Starobinsky
		inflationary model}.
	\newblock {\em Eur. Phys. J. C.}, 80:458, 2020.
	
	\bibitem{cheong:2020}
	{D. Y. Cheong, H. M. Lee and S. C. Park}.
	\newblock {Beyond the Starobinsky model for inflation}.
	\newblock {\em Phys. Lett. B}, 805:135453, 2020.
	
	\bibitem{fomin:2020}
	{I. V. Fomin, S, V, Chervon, and A. V, Tsyganov}.
	\newblock {Generalized scalar-tensor theroy of gravity reconstruction from
		physical potentiasl of a scalar field}.
	\newblock {\em Eur. Phys. J. C.}, 80:350, 2020.
	
	\bibitem{liu:2018}
	{Lei-Hua Liu}.
	\newblock {Analysis of $R^p$ inflationary model as $p \geq 2$}.
	\newblock {\em arXiv:1807.00666v3}, 2018.
	
	\bibitem{chakravarty:2015}
	{G. K. Chakravarty and S. Mohanty}.
	\newblock {Power law Starobinsky model of inflation from no-scale SUGRA}.
	\newblock {\em Phys. Lett. B}, 746:242, 2015.
	
	\bibitem{motohashi:2015}
	{H. Motohashi}.
	\newblock {Consistency relation for $R^p$ inflation}.
	\newblock {\em Phy. Rev. D}, 91:064016, 2015.
	
	\bibitem{rojas:2012}
	{Clara Rojas and V\'ictor M. Villalba}.
	\newblock {Computation of the power spectrum in chaotic
		$\frac{1}{4}\lambda\phi^4$ inflation}.
	\newblock {\em JCAP}, 003:1, 2012.
	
	\bibitem{rojas:2009}
	{Clara Rojas and V\'ictor M. Villalba}.
	\newblock {Computation of inflationary cosmological perturbations in chaotic
		inflationary scenarios using the phase-integral method}.
	\newblock {\em Phys. Rev. D}, 79:103502, 2009.
	
	\bibitem{rojas:2007c}
	{V\'ictor M. Villalba and Clara Rojas}.
	\newblock {Applications of the phase integral method ins ome inflationary
		scenarios}.
	\newblock {\em J. Phys. Conf. Ser.}, 66:012034, 2007.
	
	\bibitem{rojas:2007b}
	{Clara Rojas and V\'ictor M. Villalba}.
	\newblock {Computation of inflationary cosmological perturbations in the
		power-law inflatioary model using the phase-integral method}.
	\newblock {\em Phys. Rev. D}, 75:063518, 2007.
	
	\bibitem{casadio:2006}
	{R. Casadio, F. Finelli, A. Kamenshchik, M. Luzzi, and G. Venturi}.
	\newblock {The method of comparison equations for cosmological perturbations}.
	\newblock {\em JCAP}, 04:011, 2006.
	
	\bibitem{casadio:2005a}
	\textnormal{R}. \textnormal{Casadio}, \textnormal{F. Finelli}, \textnormal{M.
		Luzzi}, and \textnormal{G. Venturi}.
	\newblock {Improved WKB analysis of cosmological perturbations}.
	\newblock {\em Phys.Rev. D}, 71(4):043517, 2005.
	
	\bibitem{casadio:2005b}
	\textnormal{R}. \textnormal{Casadio}, \textnormal{F. Finelli}, \textnormal{M.
		Luzzi}, and \textnormal{G. Venturi}.
	\newblock {Improved WKB analysis of slow-roll inflation}.
	\newblock {\em Phys.Rev. D}, 72(10):103516, 2005.
	
	\bibitem{casadio:2005c}
	\textnormal{R}. \textnormal{Casadio}, \textnormal{F. Finelli}, \textnormal{M.
		Luzzi}, and \textnormal{G. Venturi}.
	\newblock {Higher order slow-roll predictions for inflation}.
	\newblock {\em Phys. Lett. B}, 625:1, 2005.
	
	\bibitem{habib:2005b}
	{S. Habib and A. Heinen and K. Heitmann and G. Jungman}.
	\newblock {Inflationary Perturbations and Precision Cosmology}.
	\newblock {\em Phys. Rev. D}, 71:043518, 2005.
	
	\bibitem{martin:2003a}
	\textnormal{ J.} \textnormal{Martin} and \textnormal{D. J. Schwarz}.
	\newblock {WKB approximation for inflationary cosmological perturbations}.
	\newblock {\em Phys.Rev. D}, 67(8):083512, 2003.
	
	\bibitem{habib:2002}
	{S. Habib and A. Heinen and K. Heitmann and G. Jungman and C. Molina-Par\'is}.
	\newblock {The Inflationary Perturbation Spectrum}.
	\newblock {\em Phys. Rev. Lett.}, 89:281301, 2002.
	
	\bibitem{zhu:2014a}
	Gerald Cleaver Klaus~Kirsten Tao~Zhu, Anzhong~Wang and Qin Sheng.
	\newblock Power spectra and spectral indices of k-inflation: High-order
	corrections.
	\newblock {\em Phys. Rev. D}, 90:103517, 2014.
	
	\bibitem{zhu:2014b}
	Gerald Cleaver Klaus~Kirsten Tao~Zhu, Anzhong~Wang and Qin Sheng.
	\newblock Gravitational quantum effects on power spectra and spectral indices
	with higher-order corrections.
	\newblock {\em Phys. Rev. D}, 90:063503.
	
	\bibitem{oikonomou:2020}
	Vasilis~K. Oikonomou.
	\newblock Unifying inflation with early and late dark energy epochs in axion
	f(r) gravity.
	\newblock {\em arXiv: Cosmology and Nongalactic Astrophysics}, 2020.
	
	\bibitem{kuiroukidis:2017}
	A.~Kuiroukidis.
	\newblock Inflationary $\alpha$-attractors and f(r)-gravity.
	\newblock {\em International Journal of Modern Physics A}, 32(25):1750152,
	2017.
	
	\bibitem{noriji:2017}
	Shin’ichi Nojiri, Sergei~D. Odintsov, and Vasilis~K. Oikonomou.
	\newblock Modified gravity theories on a nutshell: Inflation, bounce and
	late-time evolution.
	\newblock {\em arXiv: General Relativity and Quantum Cosmology}, 2017.
	
	\bibitem{kazuharu:2014a}
	Kazuharu Bamba, R.~Myrzakulov, S.~D. Odintsov, and L.~Sebastiani.
	\newblock {Trace-anomaly driven inflation in modified gravity and the BICEP2
		result}.
	\newblock {\em Phys. Rev. D}, 90, 2014.
	
	\bibitem{kazuharu:2014b}
	Kazuharu Bamba, R.~Myrzakulov, S.~D. Odintsov, and L.~Sebastiani.
	\newblock Trace-anomaly driven inflation in modified gravity and the bicep2
	result.
	\newblock {\em Phys. Rev. D}, 90:043505, 2014.
	
	\bibitem{kehagias:2014}
	{A. Kehagias, A. M. Dizgah, and A. Riotto}.
	\newblock {Remarks on the Starobinsky model of inflation and its descendants}.
	\newblock {\em Phys. Rev. D}, 89:043527, 2014.
	
	\bibitem{nojiri:2014}
	Shin'ichi Nojiri and Sergei~D. Odintsov.
	\newblock Mimetic f(r) gravity: Inflation, dark energy and bounce.
	\newblock {\em Modern Physics Letters A}, 29(40):1450211, 2014.
	
	\bibitem{nojiri:2003}
	Shin'ichi Nojiri and Sergei~D. Odintsov.
	\newblock {Modified gravity with negative and positive powers of the curvature:
		Unification of the inflation and of the cosmic acceleration}.
	\newblock {\em Phys. Rev. D}, 68:123512, 2003.
	
	\bibitem{noriji:2011}
	Shin'ichi Nojiri and Sergei~D. Odintsov.
	\newblock Unified cosmic history in modified gravity: From f(r) theory to
	lorentz non-invariant models.
	\newblock {\em Physics Reports}, 505(2):59--144, 2011.
	
	\bibitem{liddle:2000}
	A.~R. Liddle and D.~H. Lyth.
	\newblock {\em {Cosmological inflation and large-scale structure}}.
	\newblock Cambridge University Press, 2000.
	
	\bibitem{stewart:2001}
	{E. D Stewart and J. Gong}.
	\newblock {The density perturbation power spectrum to second-order corrections
		in the slow-roll expansion}.
	\newblock {\em Phys. Lett. B}, 510:1, 2001.
	
	\bibitem{berry:1972}
	{M. Berry and K. E. MounT}.
	\newblock {Semiclassical Approximations in Wave Mechanics}.
	\newblock {\em Rep. Prog. Phys.}, 35:315, 1972.
	
	\bibitem{abramowitz:1965}
	M.~Abramowitz and I.~A. Stegun.
	\newblock {\em Handbook of Mathematical Functions}.
	\newblock Dover, New York, 1965.
	
	\bibitem{froman:1965}
	N.~Fr\"oman and P.~O. F\"oman.
	\newblock {\em JWKB Approximation. Contribution to the Theory}.
	\newblock North-Holland, Amsterdam, 1965.
	
	\bibitem{froman:1966B}
	N.~Fr\"oman.
	\newblock Detailed analysis of some properties of the jwkb-approximation.
	\newblock {\em Ark. Fys.}, 31:381, 1966.
	
	\bibitem{froman:1974A}
	N.~Fr\"oman and P.~O. F\"oman.
	\newblock A direct method for modifying certain phase-integral approximations
	of arbitrary order.
	\newblock {\em Ann. Phys.}, 83:103, 1974.
	
	\bibitem{froman:1996}
	N.~Fr\"oman and P.~O. F\"oman.
	\newblock {\em {Phase-Integral Method. Allowing Nearlying Transition Point}},
	volume~40.
	\newblock Springer Tracts in Natural Philosophy, 1996.
	
	\bibitem{campbell:1972}
	J.~A. Campbell.
	\newblock Computation of a class of functions useful in the phase-integral
	approximation. \textnormal{I}. \textnormal{Results}.
	\newblock {\em J. Comp. Phys.}, 10:308, 1972.
	
	\bibitem{froman:2002}
	N.~Fr\"oman and P.~O. F\"oman.
	\newblock {\em Physical Problems Solved by the Phase-Integral Method}.
	\newblock Cambridge University Press, 2002.
	
	\bibitem{froman:1970A}
	N.~Fr\"oman.
	\newblock Connection formulas for certain higher order phase-integral
	approximations.
	\newblock {\em Ann. Phys.}, 61:451, 1970.
	
	\bibitem{akrami:2018a}
	Y.~Akrami \textit{et al.}
	\newblock {Planck 2018 results. I. Overview and the cosmological legacy of
		Planck}.
	\newblock {\em arXiv:1807.06205}, 2018.
	
	\bibitem{gnuplot}
	Williams and Kelley.
	\newblock {Gnuplot 4.5: an interactive plotting program}.
	\newblock 2011.
	
\end{thebibliography}


\end{document}